\documentstyle[epsfig,aps,pre]{revtex}
\def\beq{\begin{equation}}
\def\eeq{\end{equation}}
\def\beqa{\begin{eqnarray}}
\def\eeqa{\end{eqnarray}}
\def\phib{\overline{\phi }}

\begin{document}
\draft
\title{Nonequilibrium Phase Transition for a Heavy Particle in a
Granular Fluid}
\author{Andr\'{e}s Santos \cite{andres} and James W. Dufty\cite{jim}}
\address{Department of Physics, University of Florida, Gainesville, FL 32611}
\date{\today}
\maketitle

\begin{abstract}
It is shown that the homogeneous cooling state (HCS) for a heavy impurity
particle in a granular fluid supports two distinct phases. The order
parameter $\phib_{\text{s}}$ is the mean square velocity of the impurity 
particle relative to that of a
fluid particle, and the control parameter $\xi^*$ is the fluid cooling rate
relative to the impurity collision rate. For $\xi^*<1$ there is a ``normal'' phase 
for which $\phib_{\text{s}}$ scales as the fluid/impurity mass ratio, just as for a
system with elastic collisions. For $\xi^*>1$ an ``ordered'' phase occurs in
which $\phib_{\text{s}}$
is finite even for vanishingly small mass ratio, representing an extreme
violation of energy equipartition. The phenomenon can be described in terms
of a Landau-like free energy for a second order phase transition. The
dynamics leading to the HCS is studied in detail using an asymptotic
analysis of the Enskog-Lorentz kinetic equation near each phase and the
critical domain. Critical slowing is observed with a divergent relaxation
time at the critical point. The stationary velocity distributions are
determined in each case, showing a crossover from Maxwellian in the normal
phase to an exponential quartic function of the velocity that is sharply
peaked about the non-zero $\phib_{\text{s}}$ for the ordered phase. It is shown that the
diffusion coefficient in the normal phase diverges at the critical point and
remains so in the ordered phase. This is interpreted as a transition from
diffusive to ballistic dynamics between the normal and ordered phases.
\end{abstract}

\section{Introduction}

\label{sec1}

A mixture of two mechanically different fluids rapidly approaches a common
equilibrium state for times larger than a mean free time. This equilibrium
state is characterized by a common temperature or, equivalently, mean square
velocities for each type of particle that differ by their mass ratio according 
to the equipartition of
energies. Recently, the corresponding state for a granular mixture was
studied using a two component system of hard spheres with inelastic
collisions \cite{GarzoHCS}. Instead of the equilibrium state, the granular
mixture attains a homogeneous cooling state (HCS) in which all time
dependence occurs through a scaling of the particles velocities by their root
mean square velocities. Although both components have a common cooling rate
due to  the inelastic collisions in the HCS, their granular temperatures are
different. In terms of their mean square velocities, this implies a
violation of the classical equipartition theorem. The extent of the
violation depends on the mechanical differences of the particles (e.g.,
mass, diameter, coefficient of restitution), and is greatest when 
the differences are large. The
quantitative predictions of the two temperatures from an Enskog-Lorentz
kinetic theory have been confirmed by Monte Carlo simulations \cite{MontaneroHCS}.

This effect also occurs for the simplest mixture of an impurity particle in
a one component fluid. The impurity ``equilibrates'' to a common HCS with
different temperatures for the impurity and fluid particles. The dynamics of
an impurity particle of mass $m_{0}$ in a granular fluid with particles of
mass $m$ has been studied for the limiting case of $m/m_{0}\ll 1$ \cite%
{Brey1}. The description was based on the Enskog-Lorentz kinetic equation
for the impurity in a dense fluid and the fluid was taken to be in its homogeneous cooling
state (HCS). As for the case of elastic collisions, the kinetic equation
reduces to a simple Fokker-Planck equation in this limit with a velocity
independent friction coefficient. The solution to this equation approaches an
HCS for the impurity particle. As expected, the kinetic temperatures of the
two types of particles (defined in terms of their mean square velocities)
are always different although their cooling rates are the same. The ratio of
impurity to fluid thermal velocities is not simply $m/m_{0}$ as for
equipartition of energy, but has a more complex mass dependence,  {according
to}
the mechanical properties of both particles and the degree of inelasticity \
in collisions. Nevertheless, the analysis requires that this mass dependence
be such that the ratio of thermal velocities should vanish for $m/m_{0}$ $%
\rightarrow 0$ just as it would for equipartition. A single parameter $\xi
^{* }$, the ratio of the cooling rate to  the impurity--fluid
particle collision rate, characterizes the domain for which the thermal
velocity ratio vanishes, $\xi ^{* }<1$. The \ predictions of the 
Fokker-Planck equation in this domain (velocity distribution, temperature ratio,
mean square displacement, diffusion coefficient) have been confirmed by both
Monte Carlo and molecular dynamics simulation \cite{Brey2}. As $\xi ^{*
}\rightarrow 1$ the diffusion coefficient calculated from this 
Fokker-Planck equation diverges. 

The objective here is to put the analysis of \
reference \cite{Brey1} in context by extending the discussion to $\xi ^{*
}\geq 1$. A preliminary report of this work has been given in reference
\cite{PRL}.
It is found that there is a qualitative change in the state of the system at $%
\xi ^{* }=1$ that is analogous to a second order phase transition. The
order parameter $\overline{\phi }_{\text{s}}$ is the ratio of thermal velocities with a
conjugate field $h$ proportional to the mass ratio. The parameter $\xi
^{* }$ is the analogue of the inverse temperature. 
The terminology
``ordered'' is used in analogy with magnetic systems where the ordered 
phase has a non-zero order parameter (magnetization) at zero external field. 
More precisely, the ordered phase here is associated with a broken symmetry 
or scaling $\lambda h\Rightarrow \lambda \overline{\phi }_{\text{s}}$ 
 which applies for $\xi ^{*}<1$  but does not hold for $\xi ^*\geq 1$.
For $\xi ^{* }<1$
the fluid is ``normal'' with $\overline{\phi }_{\text{s}}=0$ at $h=0,$ as in the case
of a system with elastic collisions. For $\xi ^{* }>1$ an ``ordered''
state with $\overline{\phi }_{\text{s}}\neq 0$ occurs at $h=0$, representing an extreme
breakdown of equipartition. Critical slowing and qualitative changes in the
velocity distribution function for the impurity particle occur near the
transition. The diffusion coefficient diverges for $\xi ^{* }\geq 1$ and
can be understood as a transition from diffusive to ballistic motion.

 In the next section three characteristic frequencies are introduced:
the cooling rate for the fluid particles, the cooling rate for the impurity,
and the impurity--fluid collision rate. A simple estimate is obtained using
a maximum entropy distribution to construct a phenomenological overview of
the HCS, its properties for $m/m_{0}\ll 1$ (or equivalently $h\ll 1)$, and
the phase transition analogy. 
In Section \ref{sec4} the diffusion
coefficient is calculated from its Green-Kubo representation using the
leading term in a cumulant expansion of the velocity autocorrelation
function \cite{Duftyeinstein}. The diffusion coefficient is expressed as a
function of the order parameter $\overline{\phi }_{\text{s}}\left(\xi ^{*
},h\right) $, and for $\xi ^{* }<1$ the results of \cite{Brey1} are
recovered. Otherwise, at the critical point and in the ordered phase, it is divergent. 
This divergence is interpreted by
reconsideration of the Green-Kubo expression for finite times,
showing a crossover from  
diffusive  behavior in the normal phase to ballistic motion in the ordered
phase. 

A more complete description is given in
Section \ref{sec3} based on an exact asymptotic analysis 
of the Enskog-Lorentz kinetic
equation for the impurity particle velocity distribution function. This
distribution function is calculated in the critical domain showing a
crossover from Maxwellian for $\xi ^{* }<1$ to an exponential quartic
function of the velocity centered about a non-zero 
value for $\xi ^{* }>1$. The functional
form of $\overline{\phi }_{\text{s}}\left( h,\xi ^{* }\right) $ and associated
critical properties are similar to those obtained in the phenomenological
overview, with no qualitative differences. 
These results are
summarized and discussed in the last section.

\section{Phenomenological Overview}

\label{sec2}

Consider a fluid of hard, smooth, inelastic spheres of mass $m$, 
diameter $\sigma $,
and fluid--fluid particle  coefficient of normal restitution $\alpha $. 
In all of the
following it is assumed that the fluid is in its HCS. Due to the inelastic
collisions among particles the mean kinetic energy decreases as a function
of time (referred to as ``cooling''). An impurity particle of mass $m_{0}$,
diameter $\sigma _{0}$, and impurity--fluid particle 
coefficient of restitution $\alpha _{0}$ is inserted in the fluid at some initial time.
There is energy transfer between the impurity and fluid particles due to
collisions and subsequently a common HCS for the fluid and impurity is
attained where all particles have the same cooling rate. In this section a
phenomenological but accurate description of this process and the HCS is
given to present the basic ideas in a simple physical context.

\subsection{Nonlinear friction coefficient}

The primary property of interest is the ratio of the mean square velocities
for the impurity and fluid  particles 
\begin{equation}
\overline{\phi }(t)=\frac{\langle v_{0}^{2}(t)\rangle }{\langle
v^{2}(t)\rangle },  \label{2.1}
\end{equation}%
where the brackets denote an average over the initial state of the fluid
plus impurity particle. This function measures the accommodation of the
impurity particle to the fluid and will be referred to in the following as
the order parameter. The cooling rates associated with the mean square
velocities are defined by 
\begin{equation}
\xi(t) =-\partial _{t}\ln \langle v^{2}(t)\rangle ,\quad \xi _{0}(t)=-\partial
_{t}\ln \langle v_{0}^{2}(t)\rangle .  \label{2.2}
\end{equation}%
For dimensionless units it is useful to define an average impurity--fluid
particle collision rate 
\begin{equation}
\nu _{c}(t)=\frac{8}{3}h\rho\pi \overline{\sigma }^{2}g_{0}\left\langle
v(t)\right\rangle ,\quad h\equiv \frac{1+\alpha _{0}}{2}\frac{m}{%
m+m_{0}} , \label{2.3}
\end{equation}%
where $\overline{\sigma }=\left( \sigma +\sigma _{0}\right) /2$ is the
average diameter, $\rho$ is the fluid density, $g_{0}$ is the pair correlation
function for the impurity particle and a fluid particle at contact, and $%
\left\langle v(t)\right\rangle $ is the average speed of a fluid particle
in the HCS. The parameter $h$ has been introduced as a measure of the mass
ratio. As a function of $h$ this form for the collision frequency is the
same as that for elastic collisions characterizing the equilibration rate. 
A dimensionless equation for $\overline{%
\phi }(t)$ now can be written in the form 
\begin{equation}
\partial _{s}\overline{\phi }=\left(\xi ^{* }-
\xi_{0}^{* }\right)\overline{\phi } ,  \label{2.4}
\end{equation}%
where the dimensionless cooling rates and dimensionless time have been
introduced as 
\begin{equation}
\xi ^{* }=\frac{\xi }{\nu _{c}},\quad \xi _{0}^{* }=
\frac{\xi_0}{\nu _{c}},\quad ds=\nu _{c}(t)dt.  \label{2.5}
\end{equation}

To proceed it is necessary to calculate $\xi ^{* }$ and $\xi _{0}^{* }$
as functions of $\overline{\phi }$. As shown in Appendix \ref{appA}, 
these are
related to averages over the pair distribution function for two fluid
particles and for a fluid and the impurity particle, respectively. This is a
formal result since the distribution functions are not known. As a
phenomenological estimate therefore, these averages are performed using a
maximum entropy ensemble parameterized by the true mean square velocities.
The qualitative accuracy of this approximation is confirmed in 
Section \ref{sec3}. The results of Appendix \ref{appA} are 
\begin{equation}
\xi ^{* }=\frac{ 1-\alpha ^{2} }{4\sqrt{2}h}\frac{g}{g_{0}}%
\left( \frac{\sigma }{\overline{\sigma }}\right) ^{2},\quad \xi
_{0}^{* }(\phib)=\left( 1+\phib \right) ^{1/2}\left( 1-h\frac{1+\phib }{\phib }%
\right)  ,\label{2.6}
\end{equation}
where $g$ is the pair correlation function for two fluid particles at contact.
This form for the cooling rate $\xi_{0}^{* }$ of the impurity is the same as that for
elastic collisions and represents the equilibration rate. The new features
of inelasticity are primarily described by $\xi ^{* }$, which is
independent of $\phib $. The equation for $\overline{\phi }(s)$ with these
approximate forms for the cooling rates is 
\begin{equation}
\left[ \partial _{s}+\overline{\gamma} ^{* }(\overline{\phi })-\xi ^{*
}\right] 
\overline{\phi }=h\overline{n}(\overline{\phi }),  \label{2.7}
\end{equation}
which results from the decomposition $\xi_0^*(\phib)=\overline{\gamma} ^{*
}(\overline{\phi })-h \overline{n} ^{* }(\overline{\phi })/\phib$
with the definitions 
\begin{equation}
\overline{\gamma} ^{* }(\overline{\phi })=\left( 1+\overline{\phi }\right) ^{1/2},%
\quad \overline{n}(\overline{\phi })=\left( 1+\overline{\phi }\right)^{3/2}.
\label{2.8}
\end{equation}
This has the same form as would be obtained from a simple Langevin or
Brownian motion model where $\overline{\gamma} ^{* }(\overline{\phi })$ is the
``friction constant'' or nonlinear impurity--fluid collision frequency and 
$h\overline{n}(\overline{\phi })$ is the noise amplitude. The solution to this equation is
a function of time and the two parameters $\xi ^{* }$ and $h.$ The
stationary solutions $\overline{\phi }_{\text{s}}\left( \xi ^{* },h\right) $
are determined from 
\begin{equation}
\overline{\phi }_{\text{s}}=h\frac{\overline{n}(\overline{\phi }_{\text{s}})}
{\overline{\gamma} ^{* }(
\overline{\phi }_{\text{s}})-\xi ^{* }} . \label{2.8a}
\end{equation}
This form shows most clearly the effect of competition between ``friction'' on
the impurity particle and fluid cooling since $\overline{\gamma} ^{* }(
\overline{\phi }_{\text{s}})>\xi ^{* }$ is required for positive, finite solutions.
This generalizes the result obtained in reference \cite{Brey1}
which is limited to $\xi ^{* }<1$ and $h\rightarrow 0$. It is easily
verified that a unique positive solution to (\ref{2.8a}) exists for all  positive $\xi
^{* }$ and $h$ and that it is linearly stable. The latter confirms that
the HCS characterized by $\overline{\phi }_{\text{s}}$ is approached for long
times for a wide class of initial conditions. The time scale for formation
of the HCS is discussed below. For elastic collisions ($\alpha=\alpha_0=1$) 
the solution is $\overline{\phi }_{\text{s}}=h/\left( 1-h\right) =m/m_{0}$ as required by
equipartition. If only the impurity--fluid particle collisions are
inelastic (i.e., $\alpha=1$, $\xi ^{* }=0$) a recent result of Martin and Piasecki is
recovered\cite{Martin}, $\overline{\phi }_{\text{s}}=h/\left( 1-h\right) =m(1+\alpha _{0})/%
\left[ 2m_{0}+m\left( 1-\alpha _{0}\right) \right]$.

\begin{figure}[tbh]
\centerline{\epsfig{file=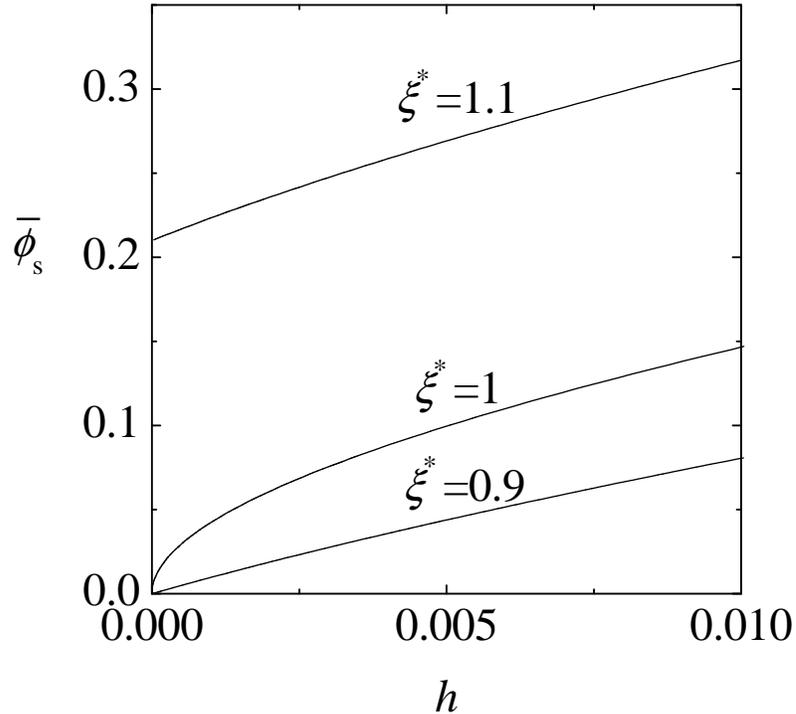,width=0.58\textwidth}}
\caption{Ratio of mean square velocities, $\phib_{\text{s}}$, as a 
function of the mass ratio parameter 
$h$ for $\xi^*=0.9$, 1, and 1.1.
\label{fig1}}
\end{figure}

\begin{figure}[tbh]
\centerline{\epsfig{file=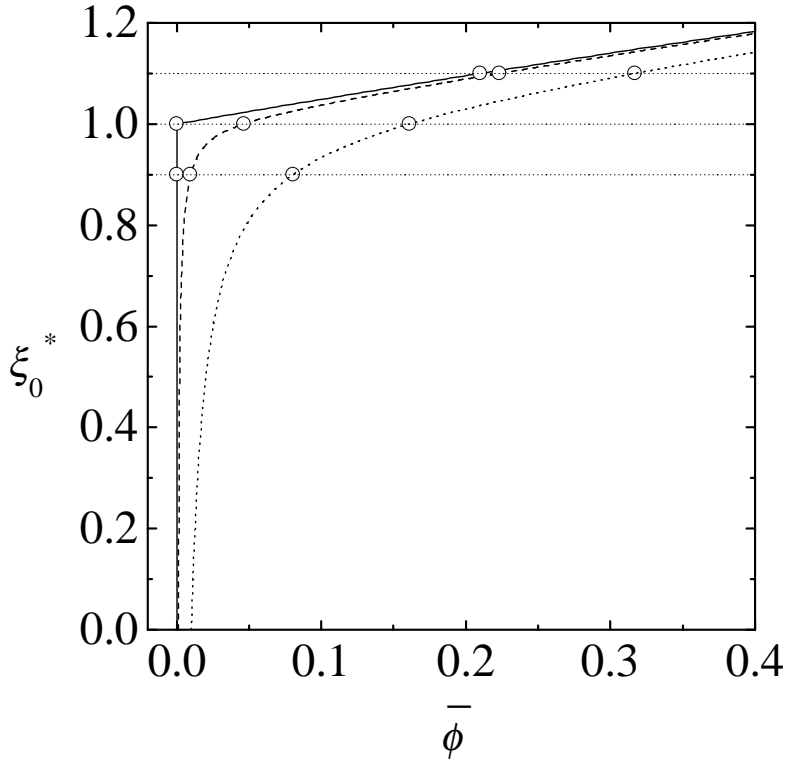,width=0.58\textwidth}}
\caption{Plot of $\xi_0^*(\phib)$, Eq.\ (\protect\ref{2.6}), for $h=10^{-2}$ 
(dotted line), $h=10^{-3}$ (dashed
line), and $h=0$ (solid line). {The intercepts of the curves with the
horizontal lines $\xi^*=0.9$, 1, and 1.1 give the corresponding values of
$\phib_{\text{s}}(\xi^*,h)$ (circles)}.
\label{fig2}}
\end{figure}

{
 More generally, Eq.\ (\ref{2.8a})  can be transformed  into 
 a cubic equation for $\phib_{\text{s}}$ whose physical solution gives $\phib_{\text{s}}(\xi^*,h)$
 for arbitrary $\xi^*$ and $h$. 
Figure \ref{fig1} shows $\overline{\phi }_{\text{s}}\left( \xi ^{* },h\right) $ as a
function of $h$ for $\xi ^{* }=0.9,1.0,$ and $1.1$.
An instructive alternative form for the determination of $\phib_{\text{s}}$ is
\beq
\xi^*=\xi_0^*(\phib_{\text{s}}).
\label{2.8ab}
\eeq
The graphical solution to Eq.\ (\ref{2.8ab}) is obtained in a plane $y$ vs
$\phib$ by finding the value of
$\phib$ at which the constant $y=\xi^*$ intercepts 
the curve $y=\xi_0^*(\phib)$, as
illustrated in Fig.\ \ref{fig2}}. 
There is seen to be a qualitative difference between the solutions for $\xi
^{* }<1$ and $\xi ^{* }>1$ in the limit of small $h$. Since
$\overline{\gamma}
^{* }({\phib })\geq \overline{\gamma}^*(0)=1$ and $\overline{n}(0)=1$ 
the asymptotic solution for $\xi ^{* }<1$ is 
\begin{equation}
\overline{\phi }_{\text{s}}\rightarrow 
h\frac{\overline{n}(0)}{\overline{\gamma}^{* }(0)-\xi ^{* }%
}=\frac{h}{1-\xi ^{* }},  \label{2.8b}
\end{equation}
which agrees 
with \cite{Brey1}. The mechanism responsible for solutions with 
$\xi ^{* }>1$ is now clear. As $\xi ^{* }$ exceeds $\overline{\gamma}^{* }(0)$
the nonlinear dependence of the friction coefficient on $\phib$ is activated to maintain
positivity of $\overline{\gamma}^{* }(\overline{\phi }_{\text{s}})-\xi ^{* }$. Since $%
\overline{\gamma}^{* }(\overline{\phi })$ is a monotonically increasing function of 
$\overline{\phi }$, positivity is possible for any choice of $\xi ^{*
}$. In general this requires that $\overline{\phi }$ must be finite even for 
$h\rightarrow 0$. This is possible if $\overline{\gamma}^{* }(\overline{\phi }_{\text{s}})
-\xi ^{* }$ is of order $h$ for small $h$ or 
$\overline{\phi }_{\text{s}}=\text{constant}+{\cal O}(h)$. 
This nonlinear dependence of the
friction coefficient on $\overline{\phi }_{\text{s}}$ provides the mechanism whereby the coupling of the
impurity particle to the fluid can be enhanced for large cooling rates: the
impurity--fluid collision frequency is increased by an increased mean
square velocity of the impurity relative to that of the fluid. This is
illustrated in Figs.\ \ref{fig1} and \ref{fig2} showing the qualitative difference between $%
\xi ^{* }<1$ and $\xi ^{* }>1$. The former admits 
$\overline{\phi }_{\text{s}}\rightarrow 0$ for $h\rightarrow 0$ whereas the latter requires $%
\overline{\phi }_{\text{s}}=$constant. In more detail, the asymptotic solution to
(\ref{2.8a}) is 
\begin{equation}
\overline{\phi }_{\text{s}}\left( \xi ^{* },h\right) \rightarrow \left\{ 
\begin{array}{ll}
({1-}\xi ^{* })^{-1}h, & \xi ^{* }<1 ,\\ 
\sqrt{2h}, & \xi ^{* }=1, \\ 
\xi ^{* 2}-1+2\xi ^{* 4}(\xi ^{* 2}-1)^{-1}h, & \xi
^{* }>1.
\end{array}
\right.  \label{2.9}
\end{equation}
The common domain of $h\rightarrow 0$  and $\xi ^{* }\rightarrow 1$ can
be obtained from (\ref{2.8a}) by the scaling $h\rightarrow \epsilon ^{2}h$, 
$\overline{\phi }_{\text{s}}\rightarrow \epsilon \overline{\phi }_{\text{s}}$, 
and $ {1-}\xi^{* } \rightarrow \epsilon \left( {1-}\xi ^{* }\right) $, 
for $\epsilon\ll 1$. The
result is the quadratic form 
\begin{equation}
h\approx (1-\xi ^{* })\overline{\phi }_{\text{s}}+
\frac{1}{2}\overline{\phi }_{\text{s}}^{2},
\label{2.9a}
\end{equation}
which has the solution 
\begin{equation}
\overline{\phi }_{\text{s}}\approx \xi ^{* }-1+ \sqrt{\left(\xi^*-1 \right)
^{2}+2h} . \label{2.9b}
\end{equation}
At $h=0$ this gives $\phib_{\text{s}}\approx \xi^*-1+|\xi^*-1|$, illustrating again the
qualitative difference between $\xi^*<1$ and $\xi^*>1$.

\begin{figure}[tbh]
\centerline{\epsfig{file=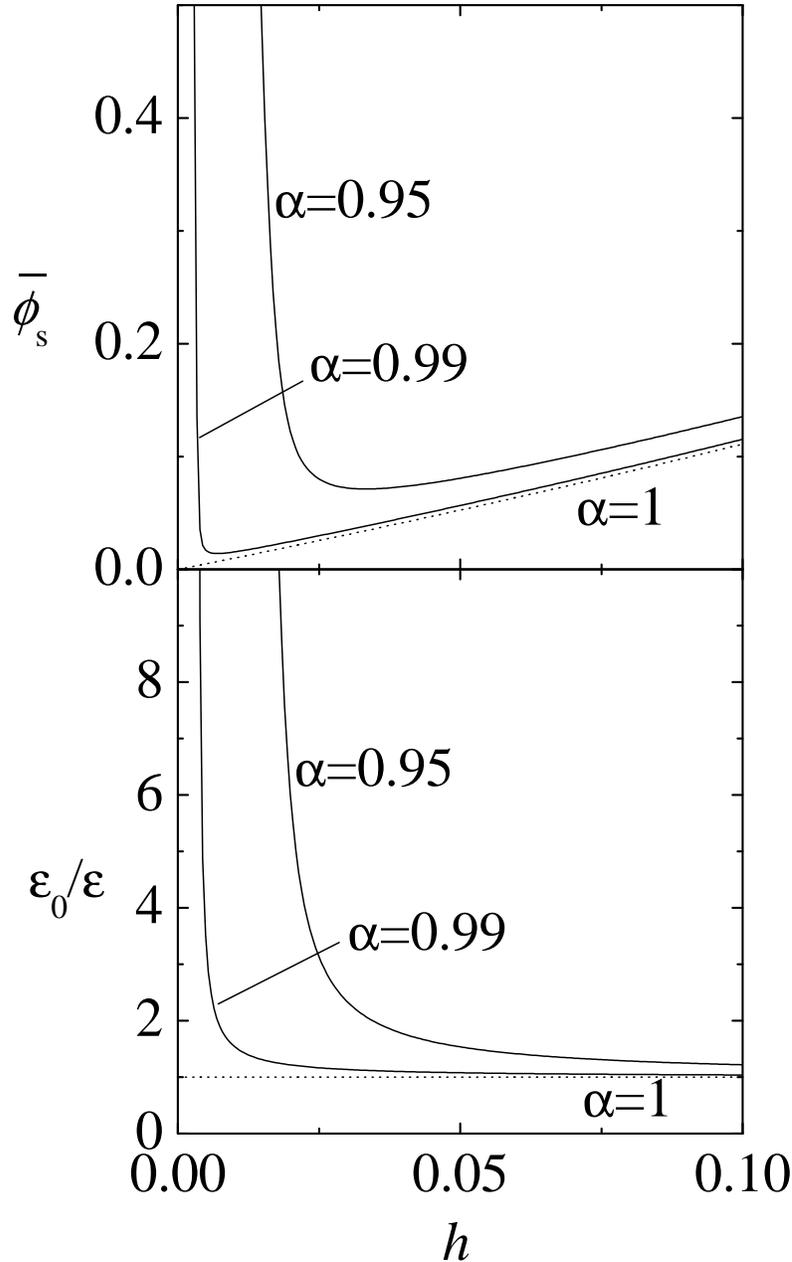,width=0.58\textwidth}}
\caption{Ratio of mean square velocities, $\phib_{\text{s}}$, and of mean
kinetic energies, $\epsilon_0/\epsilon$, as functions of the mass ratio parameter 
$h$ for $\alpha=1$, 0.95, and 0.99.
\label{fig3}}
\end{figure}

Figures \ref{fig1} and \ref{fig2} show that the fluid 
cooling rate relative to
the impurity--fluid collision rate $\xi^*$ is a control parameter
distinguishing different dependencies of $\phib_{\text{s}}$ on $h$ for small
$h$. This will be exploited in the next subsection, where $\xi^*=1$ identifies a
critical point. 
Since $\xi^*\propto (1-\alpha^2)/h$, Eq.\ (\ref{2.6}), the plots of
$\phib_{\text{s}}$ at constant $\xi^*$ require the change of both the fluid
coefficient of restitution $\alpha$ and the mass ratio parameter $h$. 
It is instructive, however, to examine the mean square velocity ratio
$\phib_{\text{s}}$ and the mean energy ratio
$\epsilon_0/\epsilon=(m_0/m)\phib_{\text{s}}$ as functions of $h$ at fixed
$\alpha<1$. In that case, $\xi^*\sim h^{-1}$ diverges in the limit $h\to 0$ and
so do $\phib_{\text{s}}\approx \xi^{*2}(1+h)^2-1\sim h^{-2}$ and
$\epsilon_0/\epsilon \sim h^{-3}$.
This is illustrated in Fig.\ \ref{fig3}, where $\phib_{\text{s}}$ and
$\epsilon_0/\epsilon$ are plotted versus $h$ for $\alpha=1$, 0.99, and 0.95
(taking, for simplicity, $\alpha_0=1$, $\sigma=\sigma_0$, $g=g_0=1$).
The dotted lines for $\alpha=1$ represent equipartition for which
$\phib_{\text{s}}\to h/(1-h)$ and $\epsilon_0/\epsilon\to 1$. For $\alpha<1$
there is a sharp deviation at sufficiently small $h$ representing the crossover
to the domain for which $\xi^*>1$.

\subsection{Representation as a phase transition}

\begin{figure}[tbh]
\centerline{\epsfig{file=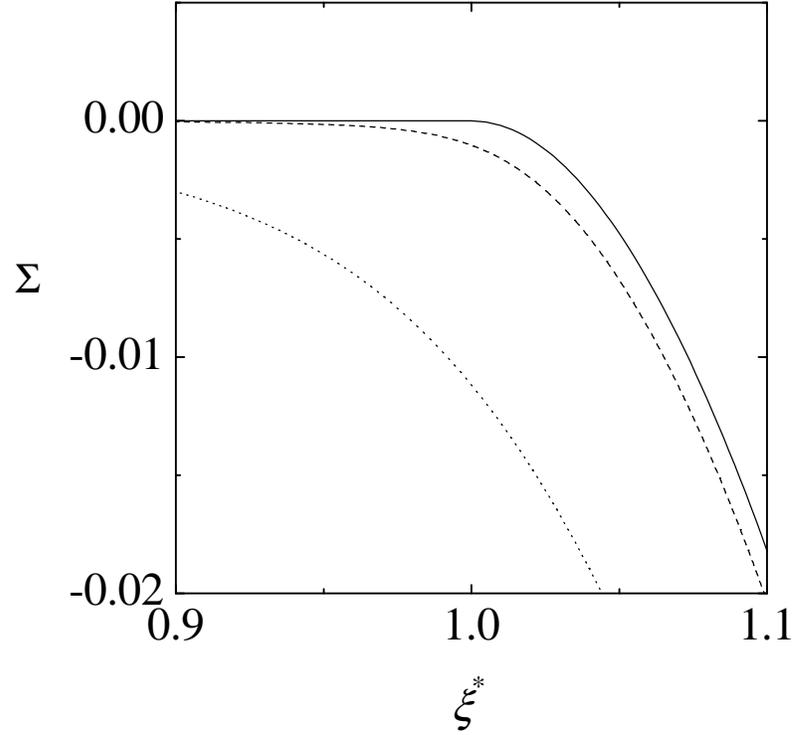,width=0.58\textwidth}}
\caption{Entropy  as a function of $\xi^*$ for $h=10^{-2}$ (dotted line), 
$h=10^{-3}$ (dashed line), and $h=0$ (solid line).
\label{fig4}}
\end{figure}

\begin{figure}[tbh]
\centerline{\epsfig{file=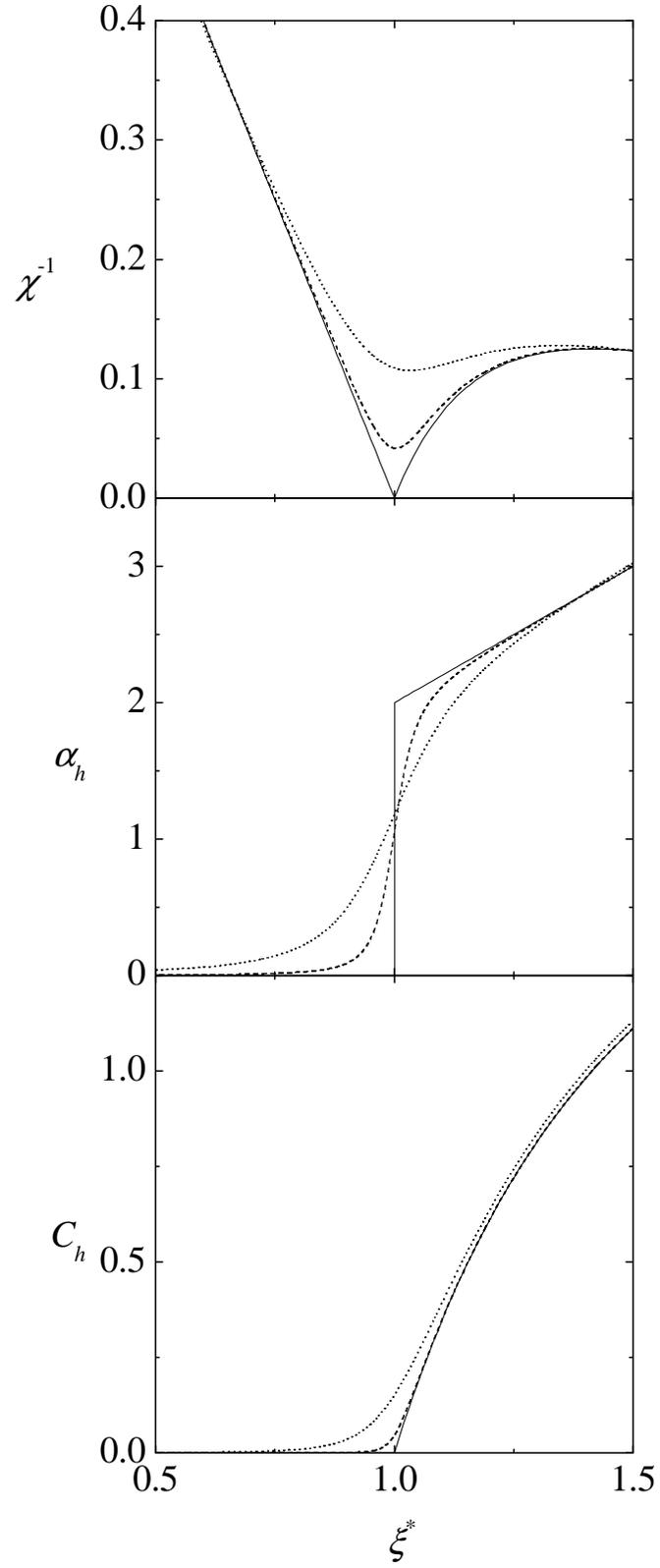,width=0.49\textwidth}}
\caption{Inverse susceptibility ($\chi^{-1}$), expansion coefficient ($\alpha_h$), 
and heat capacity ($C_h$) as functions of $\xi^*$ for $h=10^{-2}$ (dotted line), 
$h=10^{-3}$ (dashed line), and $h=0$ (solid line).
\label{fig5}}
\end{figure}

Figure \ref{fig1} and Eqs.\ (\ref{2.9}) and (\ref{2.9b}) are reminiscent of the
thermodynamics for magnetization as a function of an external magnetic
field. Below some critical temperature the magnetization is finite at zero
field,  while above that temperature it vanishes at zero field. To pursue
this analogy, consider $\overline{\phi }_{\text{s}}\left( \xi ^{* },h\right) $
as the order parameter (magnetization), $h$ the conjugate field (magnetic
field), and $\xi^*$ as the control parameter (inverse temperature).
The ``equation of state'' for the system is obtained from (\ref{2.8a}) by
solving for $h\left( \overline{\phi }_{\text{s}},\xi ^{* }\right) $,
\begin{equation}
h\left( \overline{\phi }_{\text{s}},\xi ^{* }\right) =
\frac{\overline{\phi }_{\text{s}}}
{\left( 1+\overline{\phi }_{\text{s}}\right) ^{3/2}}
\left[ \left( 1+
\overline{\phi }_{\text{s}}\right) ^{1/2}-\xi ^{* }\right] .  \label{2.10}
\end{equation}
A Helmholtz free energy can be defined in the usual way 
\begin{equation}
F\left( \overline{\phi }_{\text{s}},\xi ^{* }\right) =
\int_{0}^{\overline{\phi }_{\text{s}}}dx\,h\left( x,\xi
^{* }\right) = \overline{\phi }_{\text{s}}-\ln \left(
1+\overline{\phi }_{\text{s}}\right) -2\xi ^{* }\left[ 
\frac{2+\overline{\phi }_{\text{s}}}{\left( 1+\overline{\phi }_{\text{s}}
\right) ^{1/2}}-2\right] . \label{2.11}
\end{equation}
Next, the Gibbs free energy is obtained from the Legendre transformation 
\begin{eqnarray}
\Phi \left( \xi ^{* },h\right) &=&F\left( \overline{\phi }_{\text{s}},\xi
^{* }\right) -h\overline{\phi }_{\text{s}}(\xi ^{* },h)  \nonumber \\
&=&\left( 1-h\right) \overline{\phi }_{\text{s}}(\xi ^{* },h)-
\ln \left[ 1+\overline{\phi }_{\text{s}}(\xi ^{* },h)\right] 
-2\xi ^{* }\left\{ \frac{2+
\overline{\phi }_{\text{s}}(\xi ^{* },h)}{\left[ 1+\overline{\phi }_{\text{s}}(\xi
^{* },h)\right] ^{1/2}}-2\right\} . \label{2.12}
\end{eqnarray}
The first and second derivatives of $\Phi \left( \xi ^{* },h\right) $
provide the order parameter $\phib_{\text{s}} $, ``entropy'' $\Sigma $,
``susceptibility'' $\chi$, ``expansion coefficient'' $\alpha _{h}$, and 
``heat capacity'' $C_{h}$. The results are
\begin{equation}
\Sigma \left( \xi ^{* },h\right) =\frac{\partial \Phi \left( \xi ^{*
},h\right) }{\partial \xi ^{* }}=-2\left[ \frac{2+\overline{\phi }_{\text{s}}}{%
\left( 1+\overline{\phi }_{\text{s}}\right) ^{1/2}}-2\right] ,  \label{2.13}
\end{equation}
\begin{equation}
\chi\left( \xi ^{* },h\right) =-\frac{\partial ^{2}\Phi \left( \xi ^{* },h\right) }{\partial h^{2}}%
=\frac{\partial \phib_{\text{s}} \left( \xi ^{* },h\right) }{\partial h}
=\frac{\left( 1+\overline{\phi }_{\text{s}}\right) ^{5/2}}{\left( 1+\overline{%
\phi }_{\text{s}}\right) ^{1/2}-\xi ^{* }\left( 1-\frac{1}{2}
\overline{\phi }_{\text{s}}\right) },  \label{2.14}
\end{equation}
\begin{equation}
\alpha_h\left( \xi ^{* },h\right) =-\frac{\partial ^{2}\Phi \left( \xi ^{* },h\right) }{\partial \xi
^{* }\partial h}
=\frac{\partial \phib_{\text{s}} \left( \xi ^{* },h\right) }{\partial \xi^*}
=\chi \frac{\overline{\phi }_{\text{s}}}{\left( 1+\overline{%
\phi }_{\text{s}}\right) ^{3/2}},  \label{2.15}
\end{equation}
\begin{equation}
C_{h}\left( \xi ^{* },h\right)=-\frac{\partial ^{2}\Phi \left( \xi ^{* },h\right) }{\partial \xi
^{* 2}}=\chi ^{-1}\alpha_h ^{2}.  \label{2.16}
\end{equation}
The values of these thermodynamic properties in the limit $h=0$ for $\xi
^{* }\neq 1$ follow directly from the asymptotic forms (\ref{2.9}) for $%
\overline{\phi }_{\text{s}}$: 
\beq
\Phi \left( \xi ^{* },h=0\right) =\left\{ 
\begin{array}{ll}
 0, & \xi ^{* }<1, \\ 
\left( \xi^* -1\right) \left( 3-\xi^*\right) -2\ln \xi^* , & \xi ^{* }>1.%
\end{array}
\right. 
\eeq

\begin{equation}
\Sigma \left( \xi ^{* },h=0\right) =\left\{ 
\begin{array}{ll}
 0, & \xi ^{* }<1, \\ 
-2 \frac{\left(\xi ^{* }-1\right)^2}{\xi ^{* }} , & \xi ^{* }>1,%
\end{array}
\right.   \label{2.17}
\end{equation}
\begin{equation}
\chi \left( \xi ^{* },h=0\right) =\frac{1}{|\xi^*-1|}\left\{ 
\begin{array}{ll}
 1, & \xi ^{* }<1, \\ 
\frac{2\xi ^{* 4}}{\xi^*+1}, & \xi ^{* }>1,
\end{array}
\right.   \label{2.18}
\end{equation}
\begin{equation}
\alpha_h \left( \xi ^{* },h=0\right) =\left\{ 
\begin{array}{ll}
 0, & \xi ^{* }<1 ,\\ 
2\xi ^{* }, & \xi ^{* }>1,
\end{array}
\right.   \label{2.19}
\end{equation}
\begin{equation}
C_{h}\left( \xi ^{* },h=0\right) =\left\{ 
\begin{array}{ll}
 0, & \xi ^{* }<1, \\ 
2\frac{\xi ^{* 2}-1}{\xi ^{* 2}}, & \xi ^{* }>1.%
\end{array}
\right.   \label{2.20}
\end{equation}
With the exception of $\chi $, all thermodynamic variables vanish for $\xi
^{* }<1$ and are finite for $\xi ^{* }>1$. All are continuous at $\xi
^{* }=1$, except for $\alpha_h $ which has a finite discontinuity. The
susceptibility diverges as $\left| \xi ^{* }-1\right| \rightarrow 0$.
Thus either the discontinuity of $\alpha_h $ or the divergence of $\chi $
characterizes a second order phase transition at $\xi ^{* }=1.$ Since the
order parameter $\overline{\phi }_{\text{s}}$ behaves qualitatively as that for a
system with elastic collisions when $\xi ^{* }<1$, this will be referred
to as the ``normal'' phase. In contrast, since $\overline{\phi }_{\text{s}}\neq 0$
for $\xi ^{* }>1$ this will be called the ``ordered'' phase. 
{The entropy function $\Sigma(\xi^*,h$) is plotted versus $\xi^*$ for
$h=10^{-2}$, $10^{-3}$, and 0 in Fig.\ \ref{fig4}.
The negative value of $\Sigma$ at $h=0$ and $\xi^*>1$ is a measure of the degree of
``order'' in the ordered phase}. 
The response
functions $\chi$, $\alpha_h $, and $C_{h}$ at $h=0$, $10^{-2}$, and $%
10^{-3}$ are shown as functions of $\xi ^{* }$ in Fig.\ \ref{fig5}.

Near the critical region ($h\ll 1$, $|\xi^* -1|\ll 1$), the free energy
adopts the Landau-like form 
\begin{equation}
\Phi (\xi ^{* },h)\approx \frac{1}{2}(1-\xi ^{* })\overline{\phi }%
_{\text{s}}^{2}+\frac{1}{6}\overline{\phi }_{\text{s}}^{3}-
h\overline{\phi}_{\text{s}},
\label{2.21}
\end{equation}
which yields the critical equation of state (\ref{2.9a}), as expected. 
It is easily verified that
the free energy and the equation of state in the critical region satisfy the
scaling relations 
\begin{equation}
\Phi (\lambda (\xi ^{* }-1),\lambda ^{a}h)=\lambda ^{b}\Phi (\xi ^{*
}-1,h),\quad \overline{\phi }_{\text{s}}(\lambda (\xi ^{* }-1),\lambda
^{a}h)=\lambda ^{b-a}\overline{\phi }_{\text{s}}(\xi ^{* }-1,h)  \label{2.23}
\end{equation}
with $a=2$ and $b=3$. These scaling relations suffice to determine the
critical exponents \cite{note2} $\hat{\delta}=a/(b-a)=2$, $\hat{\beta}=b-a=1$, 
and $\hat{
\gamma}=2a-b=1$, while the critical exponent $\hat{\alpha}=2-b=-1$ is
negative, indicating that $C_{h}$ is continuous at the critical point.

\subsection{Critical dynamics}

If the ratio between the initial mean square velocities of the fluid and
impurity particles is not that given by the solution to (\ref{2.8a}), there
is an evolution to the HCS described by (\ref{2.7}) which can be written in
the Ginzburg-Landau form 
\begin{equation}
\partial _{s}\overline{\phi }=-\overline{n} (\overline{\phi })\frac{\partial \Phi
\left( \xi ^{* },h;\overline{\phi }\right) }{\partial \overline{\phi }}.
\label{2.24}
\end{equation}%
Here, $\Phi \left( \xi^*,h;\overline{\phi }\right) $ is a {\em variational\/}
 free energy given by Eq.\ (\ref{2.12}) with the order parameter $\overline{%
\phi }$ considered as an independent variable, and the kinetic coefficient is
$\overline{n}(\overline{\phi })$. The stationary solution occurs for 
$\partial \Phi \left( \xi^* ,h;\overline{
\phi }\right) /\partial \overline{\phi }=0$, which is just 
Eq.\ (\ref{2.8a}). 
It follows directly from (\ref{2.12}) and (\ref{2.24}) that $\Phi \left(
\xi^*,h;\overline{\phi }\right) $ has the properties
\beq
\Phi \left( \xi^*,h;\overline{\phi }\right) \geq \Phi \left(
\xi^*,h;\overline{\phi }_{\text{s}}\right), \quad
\partial_s \Phi \left( \xi^*,h;\overline{\phi }\right)=  -\overline{n} 
(\overline{\phi })\left[\frac{\partial \Phi
\left( \xi ^{* },h;\overline{\phi }\right) }{\partial \overline{\phi
}}\right]^2\leq 0.
\label{2.24b}
\eeq
This shows that $\Phi \left(\xi^*,h;\overline{\phi }\right) $ is a Lyapunov
function for the dynamics: it is bounded from below by the HCS solution and
monotonically approaches this bound. Consequently, the HCS solution results in
both phases for a wide class of homogeneous initial conditions and is stable.

\begin{figure}[tbh]
\centerline{\epsfig{file=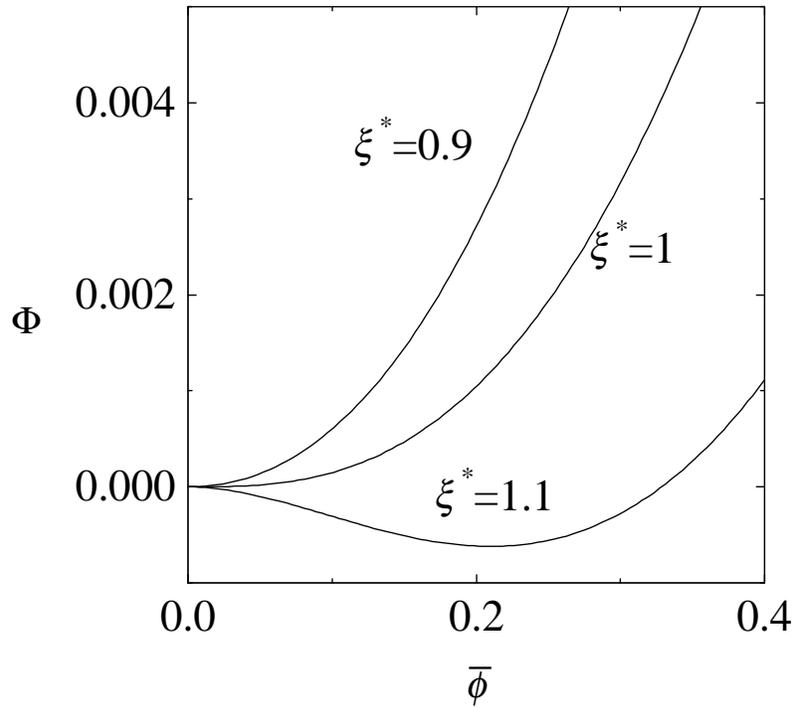,width=0.58\textwidth}}
\caption{Variational free energy $\Phi(\xi^*,h=0;\phib)$ for  $\xi^*=0.9$, 1, 
and 1.1.
\label{fig6}}
\end{figure}

\begin{figure}[tbh]
\centerline{\epsfig{file=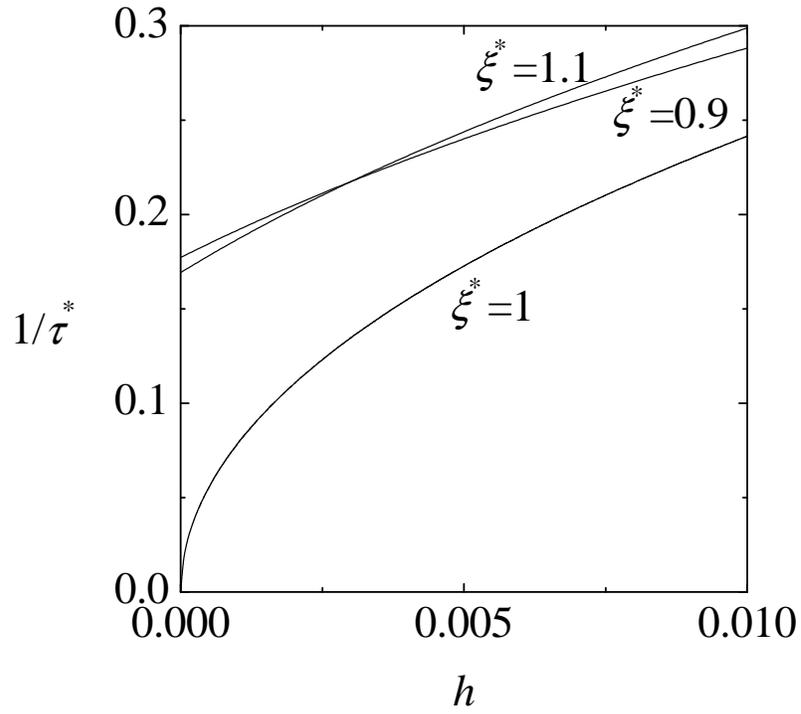,width=0.58\textwidth}}
\caption{Inverse characteristic time ${\tau^*}^{-1}$ as a function of the mass ratio parameter 
$h$ for $\xi^*=0.9$, 1, and 1.1.
\label{fig7}}
\end{figure}

{The free energy $\Phi \left( \xi^* ,h=0;\phib\right)$ is shown in Fig.\
\ref{fig6} for $\xi^*=0.9$, 1, and 1.1. As expected, the minimum is located at
$\phib=0$ for $\xi^*\leq 1$ and at $\phib\neq 0$ for $\xi^*>1$}.
For states near the HCS the evolution equation (\ref{2.24}) can be linearized and a
characteristic response time $\tau^* $ identified according to 
\begin{equation}
-\partial _{s}\ln |\overline{\phi }-\overline{\phi }_{\text{s}}|=\tau ^{*
-1}=\left( \overline{n} \chi ^{-1}\right) _{\phib_{\text{s}}}.  \label{2.26}
\end{equation}%
In the elastic limit $\tau ^{* }$ is just the equilibration time (in
terms of the number of impurity--fluid particle collisions) for the impurity
particle to attain a mean kinetic energy equal to that of the fluid
particles. Similarly, for inelastic collisions it is the time for the
impurity particle to reach a cooling rate equal to that of the fluid. This
characteristic time is a smooth function of $h$ and $\xi ^{* }$ except in
the limit $h\rightarrow 0$ where $\tau ^{* }$ diverges at $\xi ^{* }=1$.
This critical slowing follows directly from the fact that $\tau ^{*
}\propto \chi$. Otherwise, the relaxation times away from $\xi
^{* }=1$ are finite and comparable for the normal and ordered states.
Figure \ref{fig7} shows the dependence of ${\tau ^{* }}^{-1}$on $h$ for $%
\xi ^{* }=0.9$, $1$, and $1.1$.

\section{Diffusion}
\label{sec4}
Diffusion of an impurity particle in the HCS has been described in general
elsewhere \cite{Duftyeinstein,Duftygranada,lutsko}. In this section the
consequences for $h\rightarrow 0$ in the two phases are explored. A
generalized diffusion equation can be obtained by extending the familiar
methods of linear response to the granular fluid, which for long wavelengths
takes the form%
\begin{equation}
\partial _{s}n^{* }({\bf r}^{* },s)-D^{* }(s)\nabla ^{2}n^{*
}({\bf r}^*,s)=0.  \label{4.1}
\end{equation}%
Here $n^{* }({\bf r}^{* },s)$ is the dimensionless probability
density to find the impurity particle at position ${\bf r}={\bf r}^*\ell $,
where $\ell=\langle v_0^2\rangle_{\text{s}}^{1/2}/\nu_c$
is an effective mean free path,  and 
$s$ is the dimensionless time of (\ref{2.5}). The time dependent
diffusion function $D^{* }(s)$ is given exactly by a Green-Kubo expression%
\begin{equation}
D^{* }(s)=\frac{1}{3\langle v_0^{*2}\rangle_{\text{s}}}\int_{0}^{s}ds^{\prime }\langle
{\bf{v}}_{0}^{*
}\left( s^{\prime }\right) \cdot {\bf v}_{0}^{* }\rangle_{\text{s}} ,
 \label{4.2}
\end{equation}%
where ${\bf v}_0^*={\bf v}_0/\sqrt{\langle v^2\rangle_{\text{s}}}$ and the
brackets denote an average over the dimensionless HCS ensemble.
A phenomenological but accurate evaluation of the velocity autocorrelation
function is given by its exact short time behavior%
\begin{equation}
\langle{\bf v}_{0}^{* }\left( s^{\prime }\right) \cdot {\bf v}_{0}^{*
}\rangle _{\text{s}}\rightarrow \langle v_{0}^{* 2}\rangle _{\text{s}}e^{-\omega
_{D}^{* }s^\prime},  \label{4.3}
\end{equation}
\begin{equation}
\omega _{D}^{* }=-\frac{1}{2}\xi_0 ^{* }-\frac{\langle\left( L^{* }
{\bf{v}}_{0}^{* }\right) \cdot {\bf{v}}_{0}^{* }\rangle _{\text{s}}}
{\langle v_{0}^{*
2}\rangle _{\text{s}}},
\label{4.4}
\end{equation}%
where $L^*$ is the dimensionless Liouville operator [cf.\ Appendix \ref{appA}].
The dimensionless frequency $\omega _{D}^{* }$ is calculated in Appendix
\ref{appA} using the same approximation as that for the cooling rates in Section
\ref{sec2}, with the result%
\begin{equation}
\omega _{D}^{* }=\frac{1}{2}\left[ \overline{\gamma} ^{* }
(\overline{\phi }_{\text{s}})-\xi_0 ^{* }\right] . \label{4.5}
\end{equation}%
For a fluid with elastic collisions the \
approximation (\ref{4.3}) coincides with that obtained from the 
Enskog-Lorentz equation in the first Sonine approximation, and is known to be
accurate even for moderately dense systems. It is assumed that a similar 
level of accuracy extends to the inelastic case as well 
\cite{lutsko}. The diffusion function $D^{* }(s)$ becomes%
\begin{equation}
D^{* }(s)=\frac{1}{3\omega _{D}^{* }}\left(
1-e^{-\omega _{D}^{* }s}\right) . \label{4.6}
\end{equation}%
The analysis of Appendix \ref{appA} shows that $\omega _{D}^{* }>0$ for all finite 
$h$. Thus for $s\gg\omega _{D}^{* -1}$ 
\begin{equation}
D^{* }(s)\rightarrow D^{* }=\frac{1}{3\omega
_{D}^{* }}  \label{4.7}
\end{equation}%
and (\ref{4.1}) becomes the usual diffusion equation with diffusion constant 
$D^{* }$. The initial transient period is the expected ``ageing''
required for applicability of hydrodynamics (diffusion).

Consider now the behavior as $h\rightarrow 0$. Using (\ref{2.8}) and 
(\ref{2.8a}) $\omega _{D}^{* }$ can be expressed entirely in terms of 
$\overline{\phi }_{\text{s}}$ and $h$ 
\begin{equation}
\omega _{D}^{* }=\frac{1}{2}h\frac{\left( 1+\overline{\phi }_{\text{s}}\right) ^{3/2}%
}{\overline{\phi }_{\text{s}}}. \label{4.8}
\end{equation}%
Using (\ref{2.9}) this frequency behaves for $h\rightarrow 0$ as 
\begin{equation}
\omega _{D}^{* }\left( \xi ^{* },h\right) \rightarrow \left\{ 
\begin{array}{ll}
\frac{1}{2}({1-}\xi ^{* }), & \xi ^{* }<1, \\ 
\frac{1}{2}\sqrt{\frac{h}{2}}, & \xi ^{* }=1, \\ 
\frac{1}{2}\frac{\xi ^{* 3}}{\xi ^{* 2}-1}h,\hspace{0.2in} & \xi
^{* }>1.%
\end{array}%
\right.   \label{4.9}
\end{equation}%
In general, \ $\omega _{D}^{* }\left( \xi ^{* },0\right) $ is finite
below the critical point, but vanishes at and above the critical point for
$h=0$. Thus
diffusion in the sense of (\ref{4.7}) occurs at $h=0$ only for $\xi ^{*
}<1$. To understand the phenomenon for $\xi ^{* }\geq 1$ note that for 
$\omega _{D}^{* }=0$ Eq.\  (\ref{4.6})  becomes
\begin{equation}
D^{* }(s)=\frac{s}{3}.  \label{4.10}
\end{equation}%
To interpret this, take the second moment of (\ref{4.1}) with respect to 
$r^{2}$ to relate $D^{* }(s)$ to the mean square displacement of the
impurity particle,
\begin{equation}
D^{* }(s)=\frac{1}{6}\partial _{s}\langle\left| {\bf{r}}^{* }(s)-%
{\bf{r}}^{* }(0)\right| ^{2}\rangle_{\text{s}}.  \label{4.11}
\end{equation}
Thus the mean square displacement behaves as
\begin{equation}
\langle\left| {\bf{r}}^{* }(s)-{\bf{r}}^{* }(0)\right|
^{2}\rangle_{\text{s}}\rightarrow
\left\{ 
\begin{array}{ll}
6D^{* }s,& \xi ^{*}<1, \\ 
s^{2},& \xi ^{* }\geq 1.
\end{array}
\right. 
 \label{4.12}
\end{equation}
This shows that the impurity is not diffusing but rather undergoing
ballistic motion at its root mean square speed if $\xi ^{* }\geq 1$.

\section{Asymptotic Kinetic Theory}

\label{sec3}

The analysis of Sections \ref{sec2} and \ref{sec4} is based on the plausible but
uncontrolled estimate of the cooling rates for the fluid and impurity
particles using a maximum entropy ensemble (Appendix \ref{appA}). Furthermore, it is limited to a
discussion of the order parameter and diffusion but does not address other properties such
as the velocity distribution itself. In this section the results of Section %
\ref{sec2} are recovered systematically and with additional detail from  the 
Enskog-Lorentz kinetic equation for the impurity particle velocity distribution
\cite{GarzoHCS,Brey1,Duftyeinstein}. The features of interest here occur for $%
h\rightarrow 0$ so only an asymptotic representation of the kinetic theory
is required. The fluid particle distribution is independent of $h$ and its
detailed form is not required for the analysis here. The asymptotic form of
the Enskog-Lorentz equation for the impurity particle distribution $f_0({\bf %
v}_0,t)$, as a functional of the fluid particle distribution, is the focus of
this section. 

\begin{figure}[tbh]
\centerline{\epsfig{file=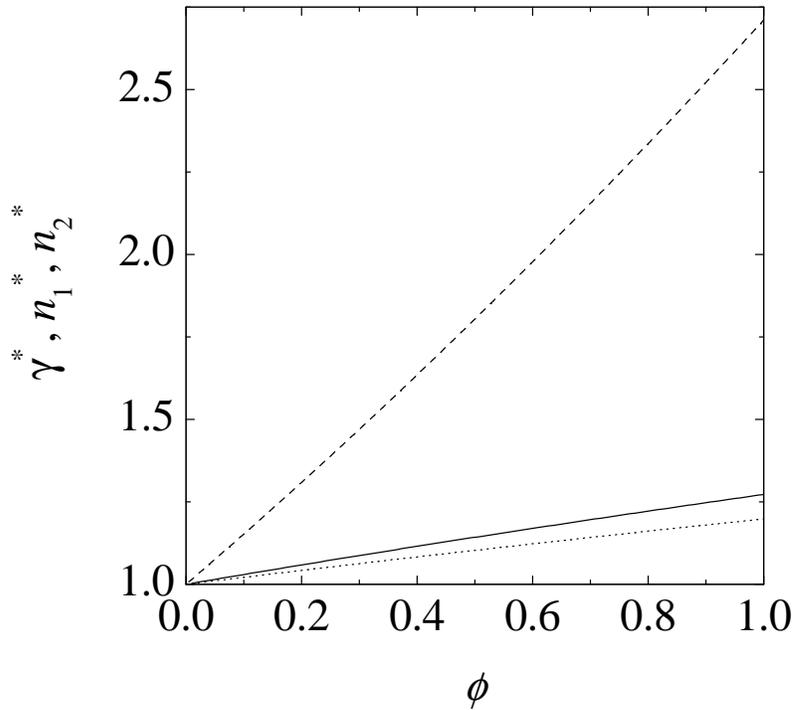,width=0.58\textwidth}}
\caption{Plot of the friction coefficient $\gamma^*(\phi)$ (solid line) and the
noise functions  $n_1^*(\phi)$ (dashed line) and
$n_2^*(\phi)$ (dotted line).
 \label{fig8}}
\end{figure}

An expansion of the impurity--fluid particle
collision operator in powers of $h$ is straightforward, leading to the
Kramers-Moyal representation \cite{vanK}. The leading terms of this expansion
have been given in  Appendix A of reference\cite{Brey1},%
\begin{eqnarray}
\partial _{t}f_0({\bf v}_0,t) &=&\frac{\partial }{\partial {\bf v}_0}\cdot \left[ h%
{\bf v}_0\gamma (v_0)f_0({\bf v}_0,t)\right] +\frac{1}{2}\frac{\partial ^{2}}{%
\partial v_{0i}\partial v_{0j}}\left\{h^2 \left[ n_{1}(v_0)\delta _{ij}\right.
\right.  \nonumber \\
&&\left. \left. +n_{2}(v_0)\left( v_{0i}v_{0j}-\frac{1}{3}\delta
_{ij}v_0^{2}\right) \right] f_0({\bf v}_0,t)\right\} +{\cal O}(h^{3}).
\label{3.1}
\end{eqnarray}%
The friction $\gamma (v_0)$ and the noise functions $n_{1}(v_0)$, $n_{2}(v_0)$ are
explicit averages over the fluid particle distribution given in Appendix
\ref{appB}.
The states of interest are functions only of the magnitude of ${\bf v}_0$.
Consequently, it is possible to introduce a variable 
\begin{equation}
\phi =\frac{v_0^{2}}{\langle v^{2}(t)\rangle }  \label{3.2}
\end{equation}%
whose average value is the order parameter $\overline{\phi }\left( s\right) $,
 where $s$ is the dimensionless time variable defined in Eq.\ (\ref{2.5}).
The distribution function for this variable is $P(\phi ,s)$, defined by 
\begin{equation}
P(\phi ,s)\equiv 4\pi f_0(v_0,t)v_0^{2}\frac{dv_0}{d\phi }=2\pi 
{\langle v^{2}(t)\rangle }^{3/2}\phi
^{1/2}f_0(v_0,t)  .\label{3.3}
\end{equation}%
Then the Kramers-Moyal expansion becomes for $P(\phi,s)$ 
\begin{eqnarray}
\partial _{s}P(\phi,s) &=&\frac{\partial }{\partial \phi }\left\{ \phi
\left[ -\xi ^{* }+\gamma ^{* }(\phi )\right] -\left( 1-\frac{2}{3}%
\frac{\partial }{\partial \phi }\phi \right) hn_{1}^{* }(\phi )+\frac{4}{5%
}\frac{\partial }{\partial \phi }\phi ^{2}hn_{2}^{* }(\phi )\right\}
P(\phi,s)  \nonumber \\
&&+{\cal O}(h^{2}) . \label{3.4}
\end{eqnarray}
Here $\xi ^{* }$ is the dimensionless cooling rate for the fluid
introduced in (\ref{2.2}). The functions $\gamma ^{* }(\phi )$, 
$n_{1}^{* }(\phi )$, and $n_{2}^{* }(\phi )$ are the dimensionless
forms of $\gamma (v_0)$, $n_{1}(v_0)$, and $n_{2}(v_0)$, respectively, given by
Eqs.\ (\ref{a27})
and (\ref{b5}) of the appendices \cite{note1}.
They are functionals of the distribution function $f(v,t)$ of the fluid in
the HCS and are normalized to have $\gamma^*(0)=1$, $n_1^*(0)=3\langle
v^3\rangle/4\langle v^2\rangle\langle v\rangle$, and $n_2^*(0)=1$. In
addition, the derivative $\gamma^{*\prime}(\phi)\equiv d\gamma^*(\phi)/d\phi$ 
at $\phi=0$ is $\gamma^{*\prime}(0)=\langle
v^{-1}\rangle\langle v^2\rangle/5\langle v\rangle$.
These functions can be accurately estimated by assuming a maximum entropy
ensemble for the fluid $f$ and
the results are given by Eqs.\ (\ref{b6a})--(\ref{b6c}). According to these estimates,  
$n_1^*(0)\simeq
1$ and $\gamma^{*\prime}(0)\simeq 3/10$. {Figure \ref{fig8} shows the friction
coefficient $\gamma^*(\phi)$ and the noise coefficients $n_1^*(\phi)$,
$n_2^*(\phi)$ according to this maximum entropy 
approximation for $f$. Hereafter, all
the plots of quantities defined in terms of those coefficients will be made
using Eqs.\ (\ref{b6a})--(\ref{b6c})}.

The Kramers-Moyal expansion is not well-ordered since the small parameter 
$h$ also multiplies the highest $\phi$ derivative. A proper asymptotic
result requires a scaling such that all higher terms in the
series are exactly zero in the appropriate limit. The simplest case is the 
{\em deterministic limit\/} for which $h= 0$ in (\ref{3.4}). 

\subsection{Deterministic limit}

If the formal limit $h= 0$ is taken in (\ref{3.4}), the equation
becomes 
\begin{equation}
\partial _{s }P_{0}(\phi,s)=\frac{\partial }{\partial \phi }
\phi \left[ -\xi ^{* }+\gamma ^{* }(\phi )\right]  P_{0}(\phi,s)
,  \label{3.5}
\end{equation}
{where the subindex 0 is used to denote quantities at $h=0$}.
The solution to this equation for sharp initial conditions $P_{0}(\phi,s=0)
=\delta \left( \phi -\phi _{0}\right) $ is 
\begin{equation}
P_{0}(\phi {\bf ,}s)=\delta \left( \phi -\overline{\phi }_{0}\left( s\right)
\right)  \label{3.6}
\end{equation}%
with 
\begin{equation}
\left[ \partial _{s}+\gamma ^{* }(\overline{\phi }_{0})-\xi ^{*
}\right] \overline{\phi }_{0}(s)=0  \label{3.7}
\end{equation}
and $\phib_0(s=0)=\phi_0$.
Thus the initial sharp distribution remains sharp and only its central value
changes in time The latter defines the macroscopic dynamics for the average value $%
\overline{\phi }_{0}(s)$. As expected, it has the form (\ref{2.7}) with a
vanishing noise. The solution for more general initial conditions
can be obtained as a superposition of the specific solution (\ref{3.6}).

\begin{figure}[tbh]
\centerline{\epsfig{file=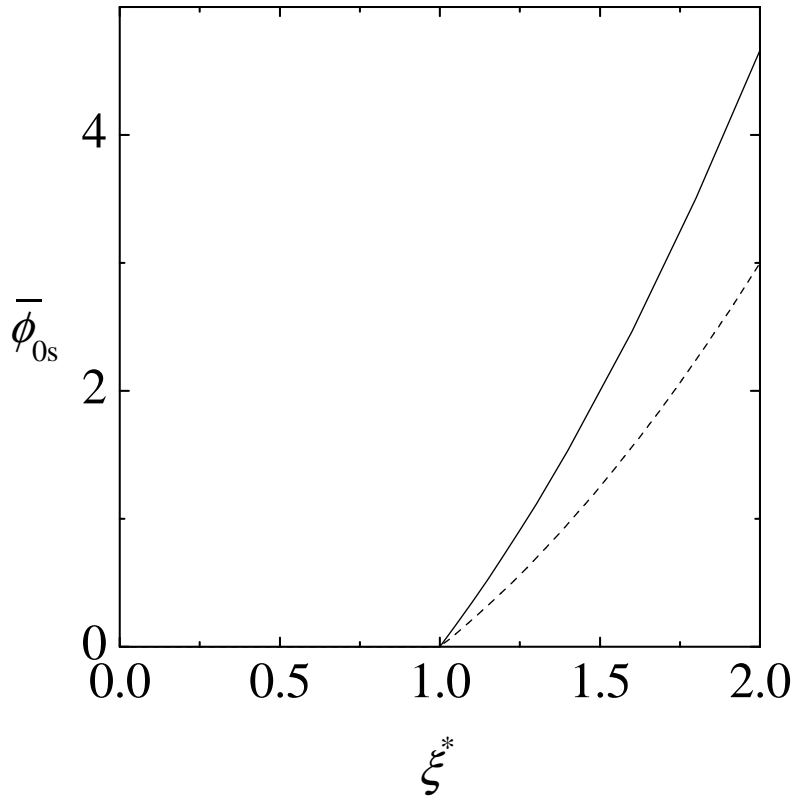,width=0.58\textwidth}}
\caption{Plot of the order parameter in the deterministic limit,
$\phib_{0\text{s}}$,
 as a function of $\xi^*$. The dashed line is the maximum entropy estimate
 $\phib_{\text{s}}={\xi^*}^2-1$ of Sec.\ \protect\ref{sec2}.
\label{fig9}}
\end{figure}

The stationary solutions are obtained from (\ref{3.7}) as the solution to 
\begin{equation}
\overline{\phi }_{0\text{s}}\left[ -\xi ^{* }+\gamma ^{* }(\overline{\phi }%
_{0\text{s}})\right] =0 . \label{3.10}
\end{equation}%
The possibilities are $\overline{\phi }_{0\text{s}}=0$ and $\gamma ^{* }(%
\overline{\phi }_{0\text{s}})=\xi ^{* }$. It is shown in Appendix \ref{appB} that 
$\gamma^{* }(0)= 1$. Therefore, the 
solution $\overline{\phi }_{0\text{s}}=0$ is stable only if $\xi ^{* }<1$. In the case $\xi ^{* }>1$
the unique stable solution is determined from $\gamma ^{* }(\overline{
\phi }_{0\text{s}})=\xi ^{* }$ with a non-zero value of $\overline{\phi
}_{0\text{s}}$.
Such solutions exist because $\gamma^{* }(\overline{\phi })$ is a monotonically
increasing function, i.e.,  $\gamma^{* }(\overline{\phi })\geq \gamma^{*
}(\overline{0})$, $\gamma^{*\prime }(\overline{\phi })\geq 0$, as proved in
Appendix \ref{appB}.
These are the two phases discussed in Section \ref{sec2}, now identified precisely
from the Enskog-Lorentz kinetic equation. The details of the ``equation of
state'' are different for this controlled analysis, but the qualitative
features of states with $\overline{\phi }_{\text{s}}=0$ and 
$\overline{\phi }_{\text{s}}\neq 0$ for $h=0$ are recovered exactly, as illustrated in Fig.\ \ref{fig9}.

\subsection{Effects of fluctuations}

A more complete description including fluctuations is obtained by a
transformation of the form $\phi =\overline{\phi }_{0}(s)+h^{p}\eta $, where 
$\overline{\phi }_{0}(s)$ is the average value of $\phi $ at $h=0$ and $\eta 
$ represents the fluctuations about this value. The power law of the scaling
for the fluctuations is determined by the requirement that the distribution
of fluctuations ${\cal P}\left( \eta ,s,h\right) =h^{p}P(\phi
,s,h)\rightarrow {\cal P}\left( \eta ,s\right) $ which is independent of $h$.
 Inverting the result in terms of $\phi $ gives the well-defined asymptotic
behavior for small $h$ \cite{vanK}. Here, attention is limited to states
near the stationary state $\overline{\phi }_{0\text{s}}$ so the chosen scaling is 
$\phi =\overline{\phi }_{0\text{s}}+h^{p}\eta $. The distribution function is
no longer sharp, as in (\ref{3.6}), but instead has a width proportional to 
$h^{p}$. The choice of $p$ is governed by the requirement that the Kramers-Moyal
 equation for the distribution of $\eta $ should truncate exactly for 
 $h=0$. The details are given in Appendix \ref{appB}, where the stationary solution in
the normal phase is found to be%
\begin{equation}
P_{\text{s}}(\phi )=\frac{3}{h\overline{\eta }_{\text{s}}}\left( \frac{3\phi }{2h\overline{%
\eta }_{\text{s}}\pi }\right) ^{1/2}e^{-3\phi /2h\overline{\eta }_{\text{s}}},\quad
\xi ^{* }<1 . \label{3.11}
\end{equation}%
and the width of the distribution is characterized by%
\begin{equation}
\overline{\eta }_{\text{s}}=\frac{n_{1}^{* }(0)}{1-\xi ^{* }}=\frac{3\langle
v^3\rangle}{4\langle v^2\rangle\langle v\rangle}\frac{1}{1-\xi^*}.  \label{3.12}
\end{equation}%
Since $\phib_{0\text{s}}=0$ in this phase, the order parameter is
$\phib_{\text{s}}=h\overline{\eta}_{\text{s}}$.
This agrees with the result of reference \cite{Brey1}, where the
distribution is recognized as a Maxwell-Boltzmann distribution for the
velocity of the impurity particle, but with a different temperature than that
of the fluid. In the present notation the impurity temperature identified
from this Maxwellian is
\begin{equation}
T_{0}=T\frac{1+\alpha_0 }{2}\overline{\eta }_{\text{s}} , \label{3.13}
\end{equation}%
where $T$ is the granular temperature of the fluid.
The phase transition is seen to occur
with a diverging kinetic temperature for the impurity particle.
If  a maximum entropy distribution is assumed for the fluid, then the right
side of (\ref{3.12}) can be evaluated to get
$\phib_{\text{s}}=h/(1-\xi^*)$, which agrees with the phenomenological theory of Section
\ref{sec2}, Eq.\ (\ref{2.9}).

\begin{figure}[tbh]
\centerline{\epsfig{file=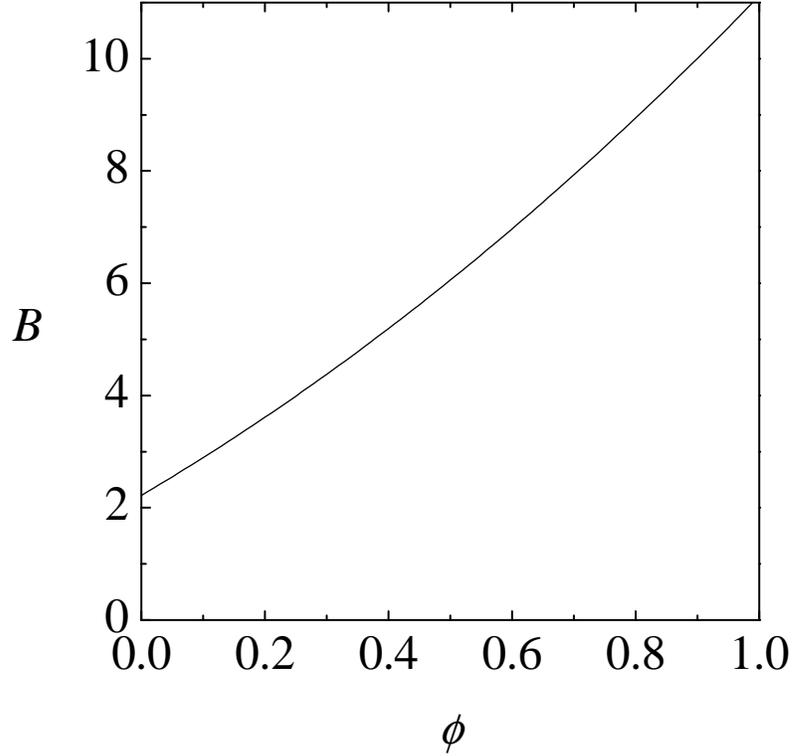,width=0.58\textwidth}}
\caption{Plot of the function $B(\phi)$ defined in Eq.\ (\protect\ref{3.15}).
 \label{fig10}}
\end{figure}

In the ordered phase a qualitatively different distribution is obtained, as
expected. It is now Gaussian in $\phi $ (quartic in velocity) and centered
about a non-zero value 
\begin{equation}
P_{\text{s}}(\phi)=\frac{1}{\sqrt{2B(\phib_{\text{s}}) h\pi }}e^{-\left( \phi
-\overline{\phi}_{\text{s}}\right)
^{2}/2B(\phib_{\text{s}})h},\quad \xi ^{* }>1,  \label{3.14}
\end{equation}%
with
\begin{equation}
B(\phi)=\frac{1}{\gamma ^{* \prime}(\phi)}\left[ \frac{2}{3}n_{1}^{* }(\phi)+\frac{4}{5}\phi 
n_{2}^{* }(\phi)\right] . 
\label{3.15}
\end{equation}%
{The function $B\left(\phi\right) $ is plotted in Fig.\ \ref{fig10}.
The width of the distribution is 
$\Delta \phi_{\text{s}}=\left[B(\phib_{\text{s}})h\right]^{1/2}$, so that as $h\to 0$ the 
distribution becomes sharply peaked about the stationary value 
$\overline{\phi }_{\text{s}}=\phib_{0\text{s}}$, where $\phib_{0\text{s}}$ is the stationary order 
parameter in the deterministic limit.
At a fixed small value of $h$ the (absolute) width $\Delta \phi_{\text{s}}$ increases,
but the  relative width
$\Delta \phi_{\text{s}}/\phib_{\text{s}}$ decreases, as
$\xi^*-1$ (and, consequently, $\phib_{\text{s}}$) increases.}

\subsection{Critical domain}

\begin{figure}[tbh]
\centerline{\epsfig{file=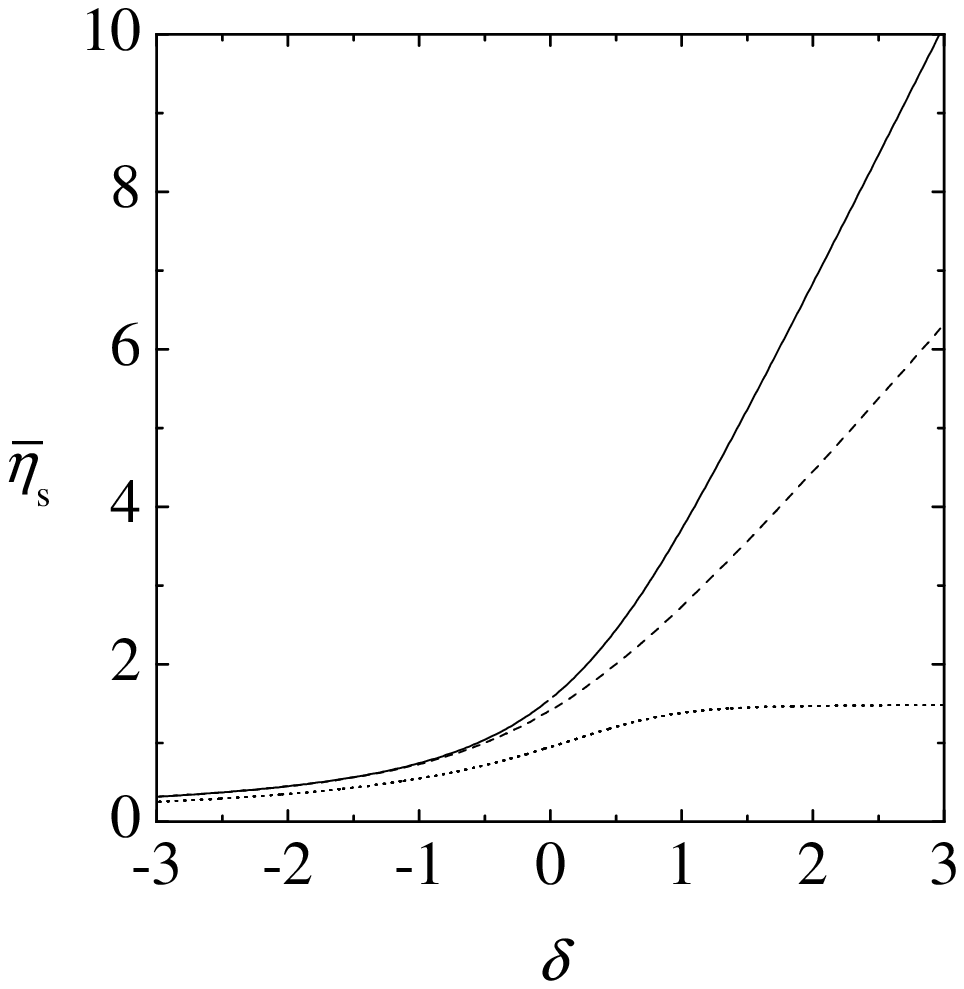,width=0.58\textwidth}}
\caption{Plot  of $\overline{\eta}_{\text{s}}(\delta)$ (solid line) and $\Delta \eta_{\text{s}}$
(dotted line). {The dashed line represents the phenomenological approximation
$\overline{\eta}_{\text{s}}(\delta)\to\delta+\sqrt{\delta^2+2}$, Eq.\ (\protect\ref{2.9b})}.
 \label{fig11}}
\end{figure}

The above results distinguish the cases of $h\rightarrow 0$ for $\xi ^{*
}<1$ and for $\xi ^{* }>1$.  A uniform description of the critical domain for
small $h$ and $\xi ^{* }\approx 1$ can be obtained by noting that $%
\overline{\phi }_{\text{s}}$ vanishes at the critical point from both phases, and
scaling the Kramers-Moyal equation according to $\xi ^{*
}-1=h^{1/2}\delta $ and $\phi =h^{1/2}\eta $. In addition a new time
variable is defined by $\tau =h^{1/2}s$. Then at $h=0$ the equation is%
\begin{equation}
\partial _{\tau}{\cal P}\left( \eta ,\tau\right) =\frac{\partial }{\partial \eta }%
\left[ -\delta \eta +\gamma ^{* \prime }(0)\eta ^{2}-\frac{1}{3}\left(
3-2\frac{\partial }{\partial \eta }\eta \right) n_{1}^{* }(0)\right]
{\cal P}\left( \eta ,\tau\right) .  \label{3.16}
\end{equation}%
The stationary distribution function is found to be
\begin{equation}
{\cal P}_{\text{s}}\left( \eta \right) =C\eta^{1/2}
\exp\left[- \frac{1}{2B(0)}\left(\eta-\frac{\delta}{\gamma ^{* \prime
}(0)}\right)^{2}\right],
\label{1.23}
\end{equation}
where $B(0)=2n_1^*(0)/3\gamma^{*\prime}(0)\simeq 2.22$ is the value at $\phi=0$ of the function defined in
Eq.\ (\ref{3.15}).
The scaled order parameter in this critical domain is then obtained from
\beq
\overline{\eta}_{\text{s}}(\delta)=\sqrt{B(0)}\frac{\int_0^\infty du\, u^{3/2}
\exp\left[-(u-2z)^{2}/2\right]}{\int_0^\infty du\, u^{1/2}
\exp\left[-(u-2z)^{2}/2\right]},
\label{n1}
\eeq
where $z\equiv \delta/2\gamma^{*\prime}(0)\sqrt{B(0)}$.
Its explicit expression is
\beq
\overline{\eta}_{\text{s}}(\delta)=\sqrt{B(0)}
\frac{\Psi_1(z)}
{4z\Psi_2(z)},
\label{n2}
\eeq
where
\beq
\Psi_1(z)=\left\{
\begin{array}{ll}
(3+4z^2)\Psi_2(z)+2z^2\left[K_{7/4}(z^2)+K_{-1/4}(z^2)-K_{5/4}(z^2)-K_{-3/4}(z^2)\right],&
z<0,\\
(3+6z^2)\Psi_2(z)+2z^2\left[I_{7/4}(z^2)+I_{-7/4}(z^2)+I_{5/4}(z^2)+I_{-5/4}(z^2)\right],&z>0,
\end{array}
\right.
\label{n3}
\eeq
\beq
\Psi_2(z)=\left\{
\begin{array}{ll}
K_{1/4}(z^2)-K_{3/4}(z^2),&z<0,\\
I_{3/4}(z^2)+I_{-3/4}(z^2)+I_{1/4}(z^2)+I_{-1/4}(z^2),& z>0,
\end{array}
\right.
\label{n4}
\eeq
$I_\nu(z)$ and $K_\nu(z)$ being the modified Bessel functions of first and
second kind, respectively. The width of the distribution
$\Delta\eta=\sqrt{\overline{\eta^2}-\overline{\eta}^2}$ in the steady state can
be obtained by taking moments in  Eq.\ (\ref{3.16}) as
\beq
\Delta\eta_{\text{s}}(\delta)=\left[\frac{n_1^*(0)+\delta\overline\eta_{\text{s}}(\delta)}
{\gamma^{*\prime}(0)}-
\overline\eta_{\text{s}}^2(\delta)\right]^{1/2}.
\label{n6}
\eeq
The asymptotic behaviors of $\overline{\eta}_{\text{s}}(\delta)$ and 
$\Delta\eta_{\text{s}}(\delta)$ are
\beq
\overline{\eta}_{\text{s}}(\delta)\to 
\left\{
\begin{array}{ll}
n_{1}^{* }(0) |\delta|^{-1},& \delta\to -\infty,\\ 
\sqrt{\lambda_1},& \delta\to 0,\\
\delta/{\gamma^{*\prime}}(0), &\delta\to \infty,
\end{array}
\right.
\label{n7}
\eeq
\beq
\Delta{\eta}_{\text{s}}(\delta)\to 
\left\{
\begin{array}{ll}
\sqrt{2/3}n_{1}^{* }(0) |\delta|^{-1},& \delta\to -\infty,\\ 
\sqrt{\lambda_2},& \delta\to 0,\\
\sqrt{B(0)}, &\delta\to \infty,
\end{array}
\right.
\label{n7.2}
\eeq
where $\lambda_1\equiv 2B(0) \left[\Gamma({5/4})/\Gamma({3/4})\right]^2\simeq
2.43$ and $\lambda_2\equiv 3B(0)/2-\lambda_1\simeq 0.90$.
In the limit $\delta\to-\infty$ we have 
$\Delta \eta_{\text{s}}/\overline{\eta}_{\text{s}}\to \sqrt{2/3}$, 
which is consistent with a Maxwell-Boltzmann distribution.
In contrast, $\Delta \eta_{\text{s}}/\overline{\eta}_{\text{s}}\to 0$ 
when $\delta\to \infty$, so that the distribution is sharp around the order 
parameter in that limit.
The dependence of the scaled order parameter 
$\overline{\eta}_{\text{s}}(\delta)$ on the scaled control parameter $\delta$
is shown in Fig.\ \ref{fig11}, where also the width of the distribution is plotted.

It is worth noting that the scaling relations $\phi=h^{1/2}\eta$, 
$\xi^*-1= h^{1/2}\delta$ are successfully captured
by the phenomenological theory of Sec.\ \ref{sec2}, except
that there the scaling function is approximated by
$\overline{\eta}_{\text{s}}(\delta)\to\delta+\sqrt{\delta^2+2}$, 
Eq.\ (\ref{2.9b}). While this function is
not quantitatively correct, especially for $\delta>0$ (cf.\ Fig.\ \ref{fig11}),
it is qualitatively consistent with the limits in
(\ref{n7}), the numerical coefficients being replaced by $\lambda_1\simeq 2.43\to 2$,
$\gamma^{*\prime}(0)\simeq 0.3\to \overline{\gamma}^{*\prime}(0)=1/2$.

Let us now go back to the unscaled variable $\phi$. 
The corresponding distribution in the critical domain is
\begin{equation}
P_{\text{s}}(\phi )\propto \phi ^{1/2}\exp \left[ -\frac{1}{2B(0)h}
\left( \phi -\frac{\xi ^{* }-1}{\gamma ^{*
\prime }(0)}\right) ^{2}\right]   \label{1.24}
\end{equation}
and the equation of state is
\beq
\phib_{\text{s}}(\xi^*,h)=h^{1/2}\overline{\eta}_{\text{s}}\left(\frac{\xi^*-1}{h^{1/2}}\right).
\label{n8}
\eeq
This result encompasses the normal and ordered phases, as well as  the
critical point.  The normal phase in the critical domain
is defined by $h\ll 1-\xi^*$ (i.e., $\delta\to -\infty$), the
ordered phase is recovered in the case $h\ll \xi^*-1$ (i.e., $\delta\to
\infty$), while the critical point corresponds to $\xi^*=1$ ($\delta=0$). Thus,
the asymptotic behaviors (\ref{n7}) translate into
\beq
\lim_{h\to 0}\overline{\phi}_{\text{s}}(\xi^*,h)= 
\left\{
\begin{array}{ll}
h n_{1}^{* }(0)/(1-\xi^*),& \xi^*\lesssim 1,\\ 
\sqrt{\lambda_1 h} ,& \xi^*=1,\\
(\xi^*-1)/{\gamma^{*\prime}}(0), &\xi^*\gtrsim 1.
\end{array}
\right.
\label{n9}
\eeq
In the normal phase, $\phib_{\text{s}}\sim h \ll 1-\xi^*$, so that the distribution
(\ref{1.24}) becomes
\begin{equation}
P_{\text{s}}(\phi )\propto \phi ^{1/2}\exp \left[ -\frac{3\phi(1-\xi^*) }{2hn_{1}^{* }(0)}
\right] ,  \label{n10}
\end{equation}
which agrees with (\ref{3.11}). {In the ordered phase, however,  the width
of the distribution is much smaller than the average value
 $\phib_{\text{s}}=(\xi^*-1)/{\gamma^{*\prime}}(0)$ [which is the solution to
$\xi^*={\gamma^{*}}(\phib_{\text{s}})$ in the critical region], so that the prefactor
$\phi^{1/2}$ in (\ref{1.24}) can
be replaced by a $\phib^{1/2}_{\text{s}}$ with the result}
\begin{equation}
P_{\text{s}}(\phi )\propto e^{-\left( \phi -\phib_{\text{s}}\right)^{2}/2B(0)h} .  \label{n11}
\end{equation}
 As expected, (\ref{n11}) agrees with (\ref{3.14})
particularized to the critical region. Finally, at the critical point the
distribution is
\beq
P_{\text{s}}(\phi )\propto \phi^{1/2}e^{-\phi^2/2B(0)h} .  \label{n12}
\end{equation}

\begin{figure}[tbh]
\centerline{\epsfig{file=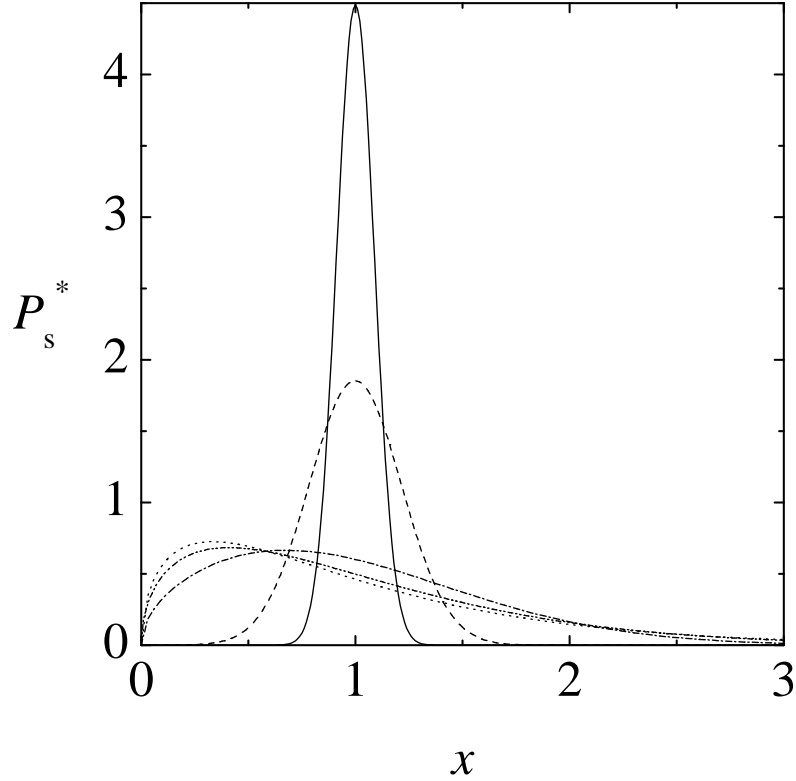,width=0.58\textwidth}}
\caption{{ Plot of the normalized distribution function
$P^{*}_{\text{s}}(x)$ for $\delta\equiv (\xi^*-1)/h^{1/2}=-\infty$ ($\cdots$),
 $-1$ (--$\cdot\cdot$--), 0 (--$\cdot$--), $2$ (-- --), and
5 (---).} \label{fig12}}
\end{figure}

As anticipated from the behavior of 
$\Delta\eta_{\text{s}}/\overline{\eta}_{\text{s}}$,
there exists a crossover in the critical domain from
the {Maxwell-Boltzmann} distribution (\ref{n10}) to the sharp distribution
(\ref{n11}) through (\ref{n12}).
Of course, the distribution function (\ref{1.24}) is more general than the
three limiting cases described by (\ref{n10})--(\ref{n12}). 
To focus on the 
{\em shape\/} of the distribution function around its average value,
define the
normalized distribution $P_{\text{s}}^*(x)=\phib_{\text{s}} P_{\text{s}}
(\phi=x\phib_{\text{s}})=\overline{\eta}_{\text{s}}{\cal
P}_{\text{s}}(\eta=x\overline{\eta}_{\text{s}})$. 
{}From Eq.\ (\ref{1.23}) we have
\beq
P_{\text{s}}^*(x)\propto x^{1/2}\exp\left[
-\frac{1}{2}\left(x\frac{\overline{\eta}_{\text{s}}}
{\sqrt{B(0)}}-2z\right)^2\right],
\quad z\equiv \delta/2\gamma^{*\prime}(0)\sqrt{B(0)}.
\label{new}
\eeq
By construction, this distribution is normalized to $\langle x\rangle =1$, 
regardless of the value of the scaled control parameter $\delta$.
The asymptotic forms of $P_{\text{s}}^*(x)$ are
\beq
P^{*}_{\text{s}}(x)\to\left\{
\begin{array}{ll}
3\sqrt{\frac{3x}{2\pi}}e^{-3x/2},& \delta\to -\infty,\\
2\frac{\left[\Gamma(5/4)\right]^{3/2}}{\left[\Gamma(3/4)\right]^{5/2}} x^{1/2}
\exp\left\{-\left[\frac{\Gamma(5/4)}{\Gamma(3/4)} x\right]^2\right\},&
\delta\to 0,\\
\sqrt{\frac{2}{\pi}}z\exp\left[-2z^2 (x-1)^2\right],& \delta\to \infty.
\end{array}
\right.
\label{n13}
\eeq
The crossover of the normalized distribution $P_{\text{s}}^*(x)$ from 
the Maxwell-Boltzmann form corresponding to $\delta\to
-\infty$ to the sharp distribution corresponding to 
$\delta=5$ is illustrated in Fig.\ \ref{fig12}.

\subsection{Critical dynamics}

To study the dynamics in the critical domain define a deviation from the
stationary solution by%
\beq
{\cal P}\left( \eta ,\tau\right) ={\cal P}_{\text{s}}\left( \eta \right) \left[
1+y\left( \eta ,\tau\right) \right] .
\label{n14}
\eeq
Equation (\ref{3.16}) becomes%
\begin{equation}
\partial _{\tau}y\left( \eta ,\tau\right) =\frac{2}{3}n_{1}^{* }(0)
\frac{1}{{\cal P}_{\text{s}}\left( \eta
\right) }\frac{\partial }{\partial \eta }{\cal P}_{\text{s}}\left( \eta \right) 
\eta \frac{\partial }{\partial \eta }y\left( \eta
,\tau\right) .
\label{n15}
\end{equation}%
{Now we define a \, ${\cal F}[y]$ by
\beq
{\cal F}[y]=\int_0^\infty d\eta\, {\cal
P}_{\text{s}}(\eta)\left[y(\eta,\tau)\right]^2.
\label{n16}
\eeq
Then, from Eq.\ (\ref{n15}) we have
\beq
\partial_\tau {\cal F}[y]=-\frac{4}{3}n_1^*(0)\int_0^\infty d\eta \,{\cal
P}_{\text{s}}(\eta) \eta \left[\frac{\partial}{\partial \eta} y(\eta,\tau)\right]^2.
\label{n17}
\eeq
Thus, ${\cal F}[y]$ has the required properties of a Lyapunov functional for the
dynamics of $y(\eta,\tau)$, namely\cite{Zubov}
\beq
{\cal F}[y]\geq 0, \quad \partial_\tau{\cal F}[y]\leq 0,
\label{n18}
\eeq
the equality being verified for $y=0$ only. This implies that for any initial condition the
solution to (\ref{n15}) evolves in time towards $y(\eta,\tau)\to 0$.}

The dynamics for  $\overline{\eta}(\tau) $ is given by%
\beq
\left( \partial _{\tau}-\delta \right) \overline{\eta }+\gamma ^{* \prime
}(0)\overline{\eta ^{2}}=n_{1}^{* }(0).
\label{n5}
\eeq
This is not a closed equation so in principle it is necessary first to solve
(\ref{3.16}) for the distribution function and then calculate
$\overline{\eta}$. However, an estimate can be obtained from (\ref{n5}) using
the approximation $\overline{\eta^2}\sim (\overline{\eta^2}_{\text{s}}
/\overline{\eta}_{\text{s}}^2)\overline{\eta}^2$. Then the linearized equation for
$x=\overline{\eta}-\overline{\eta}_{\text{s}}$ obtained from (\ref{n5}) for small $x$ is
\beq
\partial_\tau x+\left[\delta+\frac{2n_1^*(0)}{\overline{\eta}_{\text{s}}}\right]x=0.
\label{n19}
\eeq
Of course, $\delta+2n_1^*(0)/\overline{\eta}_{\text{s}} >0$ for $\delta\geq 0$. It can be
verified from Eqs.\ (\ref{n3}) and (\ref{n4}) that $\Psi_1(z)/\Psi_2(z)\geq -3$ for $z<0$, so that
$\delta+2n_1^*(0)/\overline{\eta}_{\text{s}}\geq |\delta|$ for $\delta<0$. This confirms
the above stability analysis. 
Equation (\ref{n19}) is consistent with (\ref{b20}) for the ordered phase and
(\ref{b25a}) for the normal phase. The finite relaxation time
$\overline{\eta}_{\text{s}}/2n_1^*(0)$ at the critical point is not in conflict with the
divergent relaxation time in  (\ref{2.26}) since the unit of time is different
(i.e., $\tau=\sqrt{h}s$).
\section{Discussion}
\label{sec5}
In detail application of statistical mechanics methods to the
model granular fluid of hard spheres with inelastic collisions exposes
important differences from normal fluids. First among these is the
replacement of the equilibrium Gibbs state with the time dependent
homogeneous cooling state. In the case of mixtures, the absence of detailed
balance in collisions leads to a 
breakdownwn of the usual equipartition
theorem for normal fluids. This is interesting (e.g., the HCS for a 
binary mixture
has two kinetic temperatures) but is perhaps not too surprising. In the case of
a single, mechanically different, impurity particle in a one component
granular fluid this effect is easily understood as a competition between the
average impurity--fluid collision rate $\nu _{c}$, responsible for
``equilibration,'' and the cooling rate for the fluid $\xi $ constantly
changing the reference state. This competition is most severe for conditions
such that the average impurity--fluid collision rate decreases at constant $%
\xi $, as occurs when the impurity mass is much larger than that of the
fluid particle. This requires that the nonlinear dependence of the actual
impurity--fluid collision rate on impurity mean square velocity is activated
to increase the true collision rate. As a consequence, the joint HCS for the
fluid and impurity is maintained but with a much  higher speed for the
impurity relative to that for the fluid particles. In the limit of infinite
impurity mass an extreme breakdown of equipartition occurs with the single
impurity particle attaining a finite fraction of the total kinetic energy.

 This peculiar feature distinguishes the conditions of $\xi /\nu_{c}=\xi ^{* }<1$, 
 where the distribution of energies is similar to that
for a normal fluid, from $\xi ^{* }>1$ where the distribution is
anomalous. A surprising feature of the description given here for these two
cases is the exact analogy to a second order phase transition in a normal
fluid. The order parameter is the ratio of impurity to fluid particle mean
square velocities, $\overline{\phi }_{\text{s}}$, the conjugate field is $h$ (a
measure of the mass ratio), and the role of the inverse temperature is the
relative cooling rate $\xi ^{* }$. To summarize the primary results
obtained here the following comments are offered:

\begin{itemize}

\item The nonlinearity of the impurity--fluid particle collision rate,
expressed by the dimensionless friction constant $\gamma ^{* }(\phi )$,
is essentially the same as that for an impurity in a normal fluid. In the
latter case $\phi \ll1$ for a heavy impurity particle and the relevant values
are $\gamma ^{* }(\phi )\approx \gamma ^{* }(0)=1$. However, when the
background fluid is cooling it is necessary that $\gamma ^{* }(\phi
)\approx \xi ^{* }$ so values of $\phi $ of order $1$ are selected when 
$\xi ^{* }>1$. The details of the mechanism by which the host fluid cools
is unimportant for this qualitative effect. In fact, even if all collisions
are elastic, the same two  phases would occur if the fluid were cooled by
an external thermostat.

\item The thermodynamic analogy originates from an ``equation of state'' $%
h=h\left( \overline{\phi }_{\text{s}},\xi ^{* }\right) $, obtained from the
``equilibration'' condition for the HCS. The phenomenological estimate in
Section \ref{sec2} and the exact asymptotic kinetic theory analysis of Section \ref{sec3}
are essentially the same. The ``Gibbs free energy'' obtained from
integrating the equation of state has a Landau-like form near the critical
point with critical exponents associated with the various first
and second derivatives. In particular, the susceptibility diverges,
indicating a second order phase  transition.

\item The approach to the HCS is stable in both phases. The dynamics is
governed by a Ginzburg-Landau equation defined in terms of the Gibbs free
energy. Near the phase transition there is critical slowing, with the
characteristic relaxation time diverging proportional to the susceptibility.
Alternatively this can be viewed as a change of time scale from $s$ to $\tau
=\sqrt{h}s$. 

\item The diffusion coefficient is finite in the normal phase but diverges
on approaching the transition. It remains divergent in the ordered phase. In
terms of the velocity autocorrelation function this is seen to be a
divergent relaxation time for the decay of correlations, and consequently,
the mean square displacement is characterized by ballistic rather than
diffusive dynamics. 

\item The HCS velocity distribution in both phases and in the critical
region is obtained from an exact asymptotic analysis of the Enskog-Lorentz
equation. In the normal phase away from the critical point it is a
Maxwellian with  a temperature different from that of the fluid. In the
ordered phase it is a quartic function of the velocity centered about a
non-zero average speed. The distribution function in the critical region
exhibits a continuous crossover between these distributions as the cooling
rate changes from $\xi ^{* }\leq 1$ to $\xi ^{* }\geq 1$. 

\end{itemize}

The most direct and controlled observation of the phenomena described
here would be via Monte Carlo simulation of the Boltzmann-Lorentz equation
or molecular dynamics simulation. The qualitative change in the distribution
function for the ordered phase  already has been seen in Monte Carlo
simulation, Fig.\ 6 of Ref.\ \cite{Brey2}. In principle, the Monte Carlo 
simulation could provide access to the longer
time behavior associated with critical dynamics and diffusion near the
critical point. Experimental conditions for real fluids are more difficult
to imagine, since a cooling medium for the impurity particle is required.
However, as
noted above, the cooling does not have to be associated with inelastic
collisions. Thus an impurity particle in a continuously and homogeneously
quenched fluid should exhibit the same phase transition.

The extreme breakdown of equipartition discussed in this paper extends to the
case of a mixture as well, where  a mole fraction $x_0$ of impurity particles 
exists instead of
just one impurity particle. In that case a phenomenological description
similar to that of Sec.\ \ref{sec2} shows that the critical value of the
control parameter $\xi^*$ in the limit
$h\to 0$ at finite $x_0/h$ is $\xi^*_c=1-(x_0/h)(1-\alpha_0^2)/4$, so 
that the HCS of the mixture is always in an
ordered state ($\xi^*_c=0$) if $x_0/h\geq 4/(1-\alpha_0^2)$. 
The details of
this case will be published elsewhere.
\acknowledgments

This research was supported by National Science Foundation grant PHY
9722133. A.S. acknowledges partial support from the 
Ministerio de
Ciencia y Tecnolog\'{\i}a  (Spain) through grant No.\ BFM2001-0718 and
 through a sabbatical grant No.\ PR2000-0117.

\appendix

\section{Cooling Rates and Collision Frequency}

\label{appA}

The cooling rates $\xi $ and $\xi _{0}$ for a fluid and the impurity
particle are defined by (\ref{2.2}), while the diffusion coefficient in
Section \ref{sec4} is expressed in terms of a related frequency $\omega _{D}$. They
can be written as 
\begin{equation}
\xi =-\frac{\partial _{t}\left\langle v^{2}(t)\right\rangle }{%
\left\langle v^{2}(t)\right\rangle },\quad \xi _{0}=-\frac{%
\partial _{t}\left\langle v_{0}^{2}(t)\right\rangle }{\left\langle
v_{0}^{2}(t)\right\rangle } , \quad \omega _{D}=-\frac{1}{2}\xi _{0}+\nu _{0} ,
 \label{a1}
\end{equation}%
where the impurity--fluid particle collision frequency is%
\begin{equation}
\nu _{0}=-\left.\frac{\partial _{t^\prime }\left\langle {\bf v}_{0}(t)\cdot {\bf v}%
_{0}(t+t^\prime )\right\rangle }{\left\langle v_{0}^{2}\left( t\right)
\right\rangle }\right| _{t^\prime =0}  .\label{a2a}
\end{equation}%
The subscript $0$ denotes the velocity for the impurity particle and the
brackets denote an average over the initial ensemble. The time
derivatives can be expressed in terms of the generator $L$ for the inelastic
hard sphere dynamics \cite{Duftygranada,Brey3}, 
\begin{equation}
\partial _{t}X(t)=LX(t) , \label{a3}
\end{equation}%
\begin{equation}
L={\bf v}_{0}\cdot {\bf \nabla }_{0}+\sum_{i=1}^{N}{\bf v}_{i}\cdot {\bf %
\nabla }_{i}+\sum_{i=1}^{N}T(i,0)+\frac{1}{2}\sum_{i=1}^{N}\sum_{j\neq
i}^{N}T(i,j).  \label{a4}
\end{equation}%
The binary collision operators for fluid--fluid and fluid--impurity pairs
are defined by 
\begin{equation}
T(i,j)=-\sigma ^{2}\int  d\Omega \ \Theta (-{\bf v}_{ij}\cdot \widehat{%
\bbox{\sigma}})({\bf v}_{ij}\cdot \widehat{\bbox{\sigma}})\delta ({\bf r}%
_{ij}-{\bbox{\sigma}})(b_{ij}-1),  \label{a5}
\end{equation}%
\begin{equation}
T(i,0)=-\overline{\sigma }^{2}\int d\Omega \ \Theta (-{\bf v}_{i0}\cdot 
\widehat{\bbox{\sigma}})({\bf v}_{i0}\cdot \widehat{\bbox{\sigma}})\delta (%
{\bf r}_{i0}-\overline{\bbox{\sigma}})\left( b_{i0}-1\right) ,  \label{a6}
\end{equation}%
where $\bbox{\sigma}=\sigma\widehat{\bbox{\sigma}}$,
$\overline{\bbox{\sigma}}=\overline{\sigma}\widehat{\bbox{\sigma}}$, 
and $b_{ij}$ and $b_{i0}$ transform
 the relative velocity for the pairs
into their scattered velocities and leave the center of mass invariant, 
\begin{equation}
b_{ij}{\bf v}_{ij}={\bf v}_{ij}-\left( 1+\alpha \right) \left( {\bf v}%
_{ij}\cdot \widehat{\bbox{\sigma}}\right) \widehat{\bbox{\sigma}},\quad b_{ij}{\bf G}_{ij}={\bf G}_{ij},  \label{a7}
\end{equation}%
\begin{equation}
b_{i0}{\bf v}_{i0}={\bf v}_{i0}-\left( 1+\alpha _{0}\right) \left( {\bf v}%
_{i0}\cdot \widehat{\bbox{\sigma}}\right) \widehat{\bbox{\sigma}},\quad b_{i0}{\bf G}_{i0}={\bf G}_{i0}.  \label{a8}
\end{equation}%
The various velocities and reduced masses are given by 
\begin{equation}
{\bf v}_{i0}={\bf v}_{i}-{\bf v}_{0},\quad {\bf G}_{i0}=\mu {\bf v}%
_{i}+\mu _{0}{\bf v}_{0},\quad \mu =\frac{m}{m+m_{0}},\quad
\mu _{0}=\frac{m_{0}}{m+m_{0}},  \label{a9}
\end{equation}%
\begin{equation}
{\bf v}_{ij}={\bf v}_{i}-{\bf v}_{j},\quad {\bf G}_{ij}=\frac{1}{2}%
\left( {\bf v}_{i}+{\bf v}_{j}\right) ,  \label{a10}
\end{equation}%
\begin{equation}
{\bf v}_{i}={\bf G}_{i0}+\mu _{0}{\bf v}_{i0},\quad {\bf v}_{0}={\bf %
G}_{i0}-\mu {\bf v}_{i0},\quad {\bf v}_{j}={\bf G}_{ij}-\frac{1}{2}%
{\bf v}_{ij}.  \label{a11}
\end{equation}%
In terms of the binary collision operators the cooling rates and collision
frequencies become 
\begin{equation}
\xi =-\frac{\left( N-1\right) \left\langle T(2,1)v_{1}^{2}\right\rangle
+\left\langle T(1,0)v_{1}^{2}\right\rangle }{\left\langle v^{2}\right\rangle 
},\quad \xi _{0}=-N\frac{\left\langle T(1,0)v_{0}^{2}\right\rangle }{%
\left\langle v_{0}^{2}\right\rangle },  \label{a12}
\end{equation}%
\begin{equation}
\nu _{0}=-N\frac{\left\langle {\bf v}_{0}\cdot T(1,0){\bf v}%
_{0}\right\rangle }{\left\langle v_{0}^{2}\right\rangle }.  \label{a13}
\end{equation}%
In the following it is assumed that terms of relative order $1/N$ can be
neglected. Substitution of the definitions for $T(2,1)$ and $T(1,0)$ leads
to 
\beq
T(2,1)v_{1}^{2} =-\sigma ^{2}\int d\Omega \ \Theta (-{\bf v}_{21}\cdot 
\widehat{\bbox{\sigma}})({\bf v}_{21}\cdot \widehat{\bbox{\sigma}})^2\delta (%
{\bf r}_{21}-\bbox{\sigma})\left[ \left( 1+\alpha \right) \left( 
 {\bf G}_{21}\cdot\widehat{\bbox{\sigma}}\right) 
-\frac{1}{4}\left( 1-\alpha ^{2}\right) \left(
{\bf v}_{21}\cdot \widehat{\bbox{\sigma}}\right)\right] , \label{a14}
\eeq
\beq
T(1,0)v_{0}^{2} =-\overline{\sigma }^{2}4h\int d\Omega \ \Theta (-{\bf v}%
_{10}\cdot \widehat{\bbox{\sigma}})({\bf v}_{10}\cdot \widehat{\bbox{\sigma}}%
)^2\delta ({\bf r}_{10}-\overline{\bbox{\sigma}})\left(h{\bf g}_{10}+{\bf v}_0\right)\cdot
\widehat{\bbox{\sigma}},  \label{a15}
\eeq
\beq
{\bf v}_{0}\cdot T(1,0){\bf v}_{0} =-\overline{\sigma }^{2}2h\int d\Omega
\ \Theta (-{\bf v}_{10}\cdot \widehat{\bbox{\sigma}})({\bf v}_{10}\cdot 
\widehat{\bbox{\sigma}})^2\delta ({\bf r}_{10}-\overline{\bbox{\sigma}}%
){\bf v}_0\cdot\widehat{\bbox{\sigma}} .  \label{16}
\eeq
Since these are all two-particle functions the averages in (\ref{a12}) and (%
\ref{a13}) can be reduced to integrals over the two-particle reduced
distribution functions $f^{(2)}$ and $f_{0}^{(2)}$ defined in terms of the $%
N $-particle distribution function $\rho _{\text{s}}$ as
\beq
f^{(2)}(x_{1},x_{2}) =V^{2}\int dx_{0}dx_{3}\ldots dx_{N}\rho _{\text{s}}\left(
\Gamma \right),
\label{a17.1}
\eeq
\beq
f_{0}^{(2)}(x_{0},x_{1}) =V^{2}\int dx_{2}\ldots dx_{N}\rho _{\text{s}}\left( \Gamma
\right) .  \label{a17.2}
\eeq
Here $V$ is the volume and $x_{i}$ denotes a point in the six-dimensional
phase space of particle $i$, i.e $x_{i}\Leftrightarrow \left\{ {\bf q}_{i},%
{\bf v}_{i}\right\} $. The frequencies then become 
\begin{equation}
\xi =\frac{1}{4}n\sigma ^{2}\left( 1-\alpha ^{2}\right) v_{f}\ \int d{\bf v}%
_{1}^{* }d{\bf v}_{2}^{* }\int  d\Omega f^{(2)* }({\bf v}%
_{1}^{* },{\bf v}_{2}^{* },{\bf r}_{21}=-{\bf \bbox{\sigma}})\Theta (%
{\bf v}_{21}^{* }\cdot \widehat{\bbox{\sigma}})({\bf v}_{21}^{* }\cdot 
\widehat{\bbox{\sigma}})^{3},  \label{a18}
\end{equation}%
\begin{eqnarray}
\xi _{0} &=&-4hn\overline{\sigma }^{2}v_{f}\frac{1}{\overline{\phi }}\int d%
{\bf v}_{0}^{* }d{\bf v}_{1}^{* }\int  d\Omega \ f_{0}^{(2)* }(%
{\bf v}_{0}^{* },{\bf v}_{1}^{* },{\bf r}_{10}=-\overline{{\bf %
\bbox{\sigma}}})\Theta ({\bf v}_{10}^{* }\cdot \widehat{\bbox{\sigma}})(%
{\bf v}_{10}^{* }\cdot \widehat{\bbox{\sigma}})^{2}  \nonumber \\
&&\times \left( h{\bf v}_{10}^{* }+{\bf v}_{0}^{* }\right) \cdot 
\widehat{\bbox{\sigma}},  \label{a19}
\end{eqnarray}
\begin{equation}
\nu _{0}=-2hn\overline{\sigma }^{2}v_{f}\frac{1}{\overline{\phi }}\int d{\bf %
v}_{0}^{* }d{\bf v}_{1}^{* }\int  d\Omega \ f_{0}^{(2)* }({\bf v}%
_{0}^{* },{\bf v}_{1}^{* },{\bf r}_{10}=-\overline{{\bf \bbox{\sigma}}}%
)\Theta ({\bf v}_{10}^{* }\cdot \widehat{\bbox{\sigma}})({\bf v}_{10}^{* }\cdot 
\widehat{\bbox{\sigma}})^{2}{\bf v}_{0}^{* }\cdot \widehat{\bbox{\sigma}}.
\label{a20}
\end{equation}%
All velocities have been scaled relative to $v_{f}=\sqrt{\langle v^{2}\rangle}$ and 
\begin{equation}
f^{(2)}=v_{f}^{-6}f^{(2)* }, \quad f_{0}^{(2)}=v_{f}^{-6}f_{0}^{(2)%
* }.  \label{a21}
\end{equation}%
The results at this point are still exact. It follows directly from these
results that $\xi $ and $\omega _{D}$
are manifestly positive.

\subsection{Neglect of velocity correlations}

If velocity correlations in the reduced distribution functions are neglected
on the precollision hemispheres \cite{Lutsko,Soto} they simplify to 
\begin{equation}
f^{(2)* }({\bf v}_{1}^{* },{\bf v}_{2}^{* },{\bf r}_{21}=-{\bf \bbox{\sigma}})
=gf^{* }\left( v_{1}^{* }\right)
f^{* }\left( v_{2}^{* }\right) ,\quad 
f_{0}^{(2)* }({\bf v}_{0}^{* },{\bf v}_{1}^{* },{\bf r}_{10}=-%
\overline{{\bf \bbox{\sigma}}})%
=g_{0}f_{0}^{* }\left( v_{0}^{* }\right) f^{* }\left( v_{1}^{*
}\right) ,  \label{a22}
\end{equation}%
where $g$ and $g_{0}$ are the fluid--fluid and fluid--impurity pair
correlation functions for particles at contact. The angular integrals can
now be performed to give 
\begin{equation}
\xi =\frac{1}{8}n\pi \sigma ^{2}v_{f}\ g\left( 1-\alpha ^{2}\right) \int d%
{\bf v}_{1}^{* }d{\bf v}_{2}^{* }f^{* }\left( v_{1}^*\right) f^{*
}\left( v_{2}^*\right) v_{21}^{* 3},  \label{a23}
\end{equation}%
\begin{equation}
\xi _{0}=\frac{8\pi \left\langle v^{* }\right\rangle }{3}hn\overline{%
\sigma }^{2}v_{f}g_{0}\frac{1}{\overline{\phi }}\left[\langle
v_{0}^{* 2}\gamma ^{* }(v_{0}^{* 2 })\rangle -h\langle
n_{1}^{* }(v_{0}^{* 2})\rangle \right] ,  \label{a24}
\end{equation}%
\begin{equation}
\nu _{0}=\frac{4\pi }{3}\left\langle v^{* }\right\rangle hn\overline{%
\sigma }^{2}v_{f}g_{0}\frac{1}{\overline{\phi }}\left\langle v_{0}^{*
2}\gamma ^{* }(v_{0}^{* 2})\right\rangle ,  \label{a25}
\end{equation}%
where the dimensionless functions $\gamma ^{* }(v_{0}^{* 2})$ and $%
n_{1}^{* }(v_{0}^{* 2})$ have been introduced for connection with the
discussion in Appendix \ref{appB}, 
\begin{equation}
\gamma ^{* }(v_{0}^{* 2})=\frac{3}{4v_{0}^*\left\langle
v^*\right\rangle }\int d{\bf v}_{1}^{* }f^{* }(v_{1}^{*
})v_{01}^{* }\widehat{{\bf v}}_{0}\cdot{\bf  v}_{01}^{* },\quad
n_{1}^{* }(v_{0}^{* 2})=\frac{3}{4\left\langle v^{*
}\right\rangle }\int d{\bf v}_{1}^{* }f^{* }(v_{1}^{*
})v_{10}^{* 3}.  \label{a27}
\end{equation}

\subsection{Maximum entropy ensemble}

The HCS distributions are not known exactly, although approximate
evaluations suggest they are close to Maxwellians. Therefore, to obtain an
estimate for the cooling rates and collision frequency the maximum entropy
(information) ensemble is assumed in this section. This is the Gaussian
whose density, momentum, and kinetic energy are constrained to have the same
values as for the HCS,
\begin{equation}
f^{*}\left( v^{* }\right) =\left( \frac{3}{2\pi }\right)
^{3/2}e^{-3v^{* 2}/2},\quad f_{0}^{* }\left(
v_{0}^{* }\right) =
\left( \frac{3}{2\pi\phib }\right)^{3/2}
e^{-3v_{0}^{* 2}/2\overline{\phi }}\;.  \label{a28}
\end{equation}
This gives the results 
\begin{equation}
\xi =\frac{4}{3}\sqrt{\frac{\pi }{3}}n\sigma ^{2}v_{f}\ g\left( 1-\alpha
^{2}\right) ,\quad \xi _{0}=2\nu _{0}\left( 1-h\frac{ 1+%
\overline{\phi } }{\overline{\phi }}\right) ,  \label{a29}
\end{equation}%
\begin{equation}
\nu _{0}=\frac{8}{3}\sqrt{\frac{2\pi }{3}}hn\overline{\sigma }%
^{2}v_{f}g_{0}\left( 1+\overline{\phi }\right) ^{1/2}.  \label{a30}
\end{equation}

\section{Asymptotic kinetic equations}

\label{appB}

The analysis here is based on the Enskog-Lorentz equation to describe the
distribution function for the impurity particle. Interest is restricted to the
case of small ratio of fluid particle mass to impurity particle mass. To
obtain an asymptotic form for the kinetic equation, first a Kramers-Moyal
expansion is performed to second order in the mass ratio. This accounts for
the dependence of the collisional changes on the mass ratio. Subsequently,
two different expansions are performed for the final asymptotic form
depending on the value of a control parameter $\xi^* $.

\subsection{Kramers-Moyal Expansion}

The Kramers-Moyal expansion of the Enskog-Lorentz equation has been
obtained in Appendix A of reference\cite{Brey1}. The result is%
\begin{eqnarray}
\partial _{t}f_0({\bf v}_0,t) &=&\frac{\partial }{\partial {\bf v}_0}\cdot \left[ h%
{\bf v}_0\gamma (v_0)f_0({\bf v}_0,t)\right] +\frac{1}{2}\frac{\partial ^{2}}{%
\partial v_{0i}\partial v_{0j}}\left\{h^2 \left[ n_{1}(v_0)\delta _{ij}\right.
\right.  \nonumber \\
&&\left. \left. +n_{2}(v_0)\left( v_{0i}v_{0j}-\frac{1}{3}\delta
_{ij}v_0^{2}\right) \right] f_0({\bf v}_0,t)\right\} +{\cal O}(h^{3}).
\label{b1}
\end{eqnarray}%
 The friction $\gamma (v_0)$
and noise $n_{1}(v_0),n_{2}(v_0)$ are%
\begin{equation}
\gamma (v_0)=\frac{\nu _{c}}{2h}\gamma ^{* }(\phi ),\quad n_{1}(v_0)=%
\frac{\nu _{c}}{3h}v_{f}^{2}n_{1}^{* }(\phi ),\quad n_{2}(v_0)=%
\frac{3\nu _{c}}{5h}n_{2}^{* }(\phi ),  \label{b2}
\end{equation}%
where $\nu _{c}=\frac{8}{3}hn\pi \overline{\sigma }^{2}g_{0}\left\langle
v\right\rangle $ is the characteristic impurity collision frequency
introduced in (\ref{2.3}). Also $\ \gamma ^{* }(\phi )$ and $n_{1}^{*
}(\phi )$ have been defined in (\ref{a27}),  and%
\begin{equation}
n_{2}^{* }(\phi )=\frac{15}{16\left\langle v^{* }\right\rangle
\phi}\int d{\bf v}_{1}^{* }f^{* }(v_{1}^{* })v_{01}^{* }\left[
\left( {\bf v}_{01}^{* }{\bf \cdot }\widehat{{\bf v}}_0\right) ^{2}-%
\frac{1}{3}v_{01}^{* 2}\right] .  \label{b5}
\end{equation}%
The dimensionless variables are 
\begin{equation}
\phi =v_0^{* 2},\quad v_0^{* }=v_0/v_{f},\quad {\bf v}_{1}^{*
}={\bf v}_{1}/v_{f},\quad {\bf v}_{01}^{* }=\phi ^{1/2}\widehat{{\bf v}}_0
-{\bf v}_{1}^{* },\quad v_{f}=\sqrt{\langle v^{2}(t)\rangle}.
\label{b6}
\end{equation}

The analysis of the deterministic limit in the text makes use of the
property $\gamma ^{* }(\phi )\geq \gamma ^{* }(0)=1$. To prove this,
first perform the angle integrations to get%
\begin{eqnarray}
\gamma ^{* }(\phi ) &=&1+\frac{1}{5\left\langle v^*\right\rangle }\left[
\phi \left\langle v^{* -1}\right\rangle -\pi \phi \int_{0}^{\sqrt{%
\phi }}dv^{* }f^{* }(v^{* })v^{* }\left( 4+\frac{%
v^{* }}{\sqrt{\phi }}\right) \left( 1-\frac{v^{* }}{\sqrt{\phi 
}}\right) ^{4}\right]  \nonumber \\
&\geq &1+\frac{1}{5\left\langle v^*\right\rangle }\left[ \phi \left\langle
v^{* -1}\right\rangle -4\pi \phi \int_{0}^{\sqrt{\phi }}dv^{*
}f^{* }(v^{* })v^{* }\right] ,  \label{b6.1}
\end{eqnarray}%
where the inequality results from $4\geq \left( 4+x\right) \left( 1-x\right)
^{4}$ for $x\leq 1$. Next, writing out the contribution from $\left\langle
v^{* -1}\right\rangle $ explicitly gives the desired result%
\begin{equation}
\gamma ^{* }(\phi )\geq 1+\frac{4\pi \phi }{5\left\langle
v^*\right\rangle }\int_{\sqrt{\phi }}^{\infty }dv^{* }f^{*
}(v^{* })v^{* }\geq \gamma ^{* }(0)=1.  \label{b6.2}
\end{equation}%
Analogously, it is possible to prove that $\gamma^{*\prime}(\phi)\geq 0$:
\beqa
\gamma^{*\prime}(\phi)&=&\frac{1}{5\left\langle v^*\right\rangle }\left[
 \left\langle v^{* -1}\right\rangle -\frac{\pi}{2} \int_{0}^{\sqrt{%
\phi }}dv^{* }f^{* }(v^{* })v^{* }\left( 8+9\frac{%
v^{* }}{\sqrt{\phi }}+3\frac{v^{*2}}{\phi}\right) \left( 1-\frac{v^{* }}{\sqrt{\phi 
}}\right) ^{3}\right]  \nonumber \\
&\geq &\frac{1}{5\left\langle v^*\right\rangle }\left[\left\langle
v^{* -1}\right\rangle -4\pi \int_{0}^{\sqrt{\phi }}dv^{*
}f^{* }(v^{* })v^{* }\right]\nonumber\\
&=& \frac{4\pi}{5\left\langle v^*\right\rangle }
\int_{\sqrt{\phi }}^{\infty }dv^{* }f^{*
}(v^{* })v^{* }\geq 0.
\label{b6.1bis}
\end{eqnarray}
The remaining analysis of the text and below does not require the explicit
forms for $\gamma ^{* }(\phi )$, $n_{1}^{* }(\phi )$, and $n_{2}^{*
}(\phi )$. However, for the illustrations in the graphs an excellent approximation
is obtained using the maximum entropy ensemble (\ref{a28}) for the fluid; no assumption is
required regarding the impurity particle distribution. The resulting
integrals can be performed with the results%
\begin{equation}
\gamma ^{* }(\phi )=\frac{1}{8\phi }\left( 1+3\phi \right) e^{-3\phi /2}-%
\frac{1}{16\phi ^{3/2}}\sqrt{\frac{2\pi }{3}}\left( 1-6\phi -9\phi
^{2}\right)\text{erf}
\left( \sqrt{3\phi /2}\right) ,  \label{b6a}
\end{equation}%
\begin{equation}
n_{1}^{* }(\phi )=\frac{1}{8 }\left( 5+3\phi \right) e^{-3\phi /2}+%
\frac{1}{16}\sqrt{\frac{2\pi }{3\phi }}\left( 3+18\phi +9\phi ^{2}\right) 
\text{erf}
\left( \sqrt{3\phi /2}\right) ,  \label{b6b}
\end{equation}%
\begin{equation}
n_{2}^{* }(\phi )=\frac{5}{48\phi ^{2}}\left( -1+2\phi +3\phi ^{2}\right)
e^{-3\phi /2}+\frac{5}{96\phi ^{5/2}}\sqrt{\frac{2\pi }{3}}\left( 1-3\phi
+9\phi ^{2}+9\phi ^{3}\right) 
\text{erf}
\left( \sqrt{3\phi /2}\right) .  \label{b6c}
\end{equation}

In the following only solutions that depend on the magnitude of ${\bf v}_0$ are
considered. Since the order parameter is the average of $\phi $,
\begin{equation}
\overline{\phi }(t)=\int d{\bf v}_0 v_0^{*2} f_0(v_0,t),  \label{b7}
\end{equation}%
it is appropriate to change variables from $v_0$ to $\phi $. In addition, the
dimensionless time scale of (\ref{2.5}) is introduced. This is accomplished
by defining the new distribution function $P(\phi ,s)$ by 
\beq
P(\phi ,s)\equiv 4\pi f_0(v_0,t)v_0^{2}\frac{dv_0}{d\phi }=2\pi
v_{f}^{3}\phi ^{1/2}f_0(v_0,t), 
\eeq
or 
\begin{equation}
f_0(v_0,t)=\frac{1}{2\pi v_{f}^{3}}\phi ^{-1/2}P(\phi  ,s).
\label{b8}
\end{equation}%
The Kramers-Moyal equation (\ref{b1}) becomes for $P(\phi ,s)$ 
\begin{eqnarray}
\partial _{s}P(\phi,s) &=&\frac{\partial }{\partial \phi }\left\{ \phi
\left[ -\xi ^{* }+\gamma ^{* }(\phi )\right] -\left( 1-\frac{2}{3}%
\frac{\partial }{\partial \phi }\phi \right) hn_{1}^{* }(\phi )+\frac{4}{5%
}\frac{\partial }{\partial \phi }\phi ^{2}hn_{2}^{* }(\phi )\right\}
P(\phi,s)  \nonumber \\
&&+{\cal O}(h^{2}).
\label{b9}
\end{eqnarray}
The deterministic limit, $h=0$, is described in the text. In the
following an outline of the fluctuations about this determistic limit is
given.

\subsection{Expansion around $\overline{\protect\phi }_{0\text{s}}$}

The effects of finite $h$ represent ``noise'' which broadens the width of
the initial delta function as the system evolves. To include such effects
consider solutions of the form 
\begin{equation}
P(\phi {\bf ,}s,h)=h^{-p}{\cal P}(\frac{\phi -\overline{\phi }_{0\text{s}}}{h^{p}}%
{\bf ,}s,h),  \label{b16}
\end{equation}%
such that the limit $\lim_{h\rightarrow 0}{\cal P}(\eta {\bf ,}s,h)={\cal P}(\eta {\bf ,}s)$ is finite and independent of $h$. 
The choice of
reference state $\overline{\phi }_{0\text{s}}$ given by (\ref{3.10}) implies
initial conditions that do not deviate too much from the stationary state.
To find such solutions, define a change of variables in (\ref{b9}) by 
\begin{equation}
\phi =\overline{\phi }_{0\text{s}}+h^{p}\eta ,\quad P(\phi
,s,h)=h^{-p}{\cal P}\left( \eta ,s,h\right) .  \label{b17}
\end{equation}%

In the ordered phase a non-trivial equation for ${\cal P}$ is obtained with
the choice $p=1/2$, 
\begin{equation}
\partial _{s}{\cal P}\left( \eta ,s,0\right) \rightarrow \frac{\partial }{%
\partial \eta }\left\{ \eta \overline{\phi }_{0\text{s}}\gamma ^{* \prime}(%
\overline{\phi }_{0\text{s}})+\overline{\phi }_{0\text{s}}\left[ \frac{2}{3}n_{1}^{* }(%
\overline{\phi }_{0\text{s}})+\frac{4}{5}\overline{\phi }_{0\text{s}}n_{2}^{* }(%
\overline{\phi }_{0\text{s}})\right] \frac{\partial }{\partial \eta }\right\}
{\cal P}\left( \eta ,s,0\right) ,\quad \xi ^{* }>1.  \label{b19}
\end{equation}%
where it has been recognized that $\xi ^{* }-\gamma ^{* }(\overline{%
\phi }_{\text{s}})=0$ in this phase. The average value of $\eta $ obeys the
equation 
\begin{equation}
\partial _{s}\overline{\eta }\left( s\right) =-\overline{\phi }
_{0\text{s}}\gamma ^{* \prime }(\overline{\phi }_{0\text{s}})\overline{\eta }\left(
s\right) ,  \label{b20}
\end{equation}%
which is the linearized form of the deterministic dynamics (\ref{3.7}) for 
$\xi ^{* }>1$. Stability is assured by  $\gamma ^{* \prime }(\overline{%
\phi }_{0\text{s}})\geq 0$, which is seen to be the case using (\ref{b6.1bis}). The
stationary solution to (\ref{b20}) is $\overline{\eta }_{\text{s}}=0$ so there are no
corrections to $\overline{\phi }_{0\text{s}}$ in this limit. The stationary
solution for the distribution function is obtained from 
\begin{equation}
\left[ \frac{2}{3}n_{1}^{* }(\overline{\phi }_{0\text{s}})+%
\frac{4}{5}\overline{\phi }_{0\text{s}}n_{2}^{* }(\overline{\phi }_{0\text{s}})\right] 
\frac{\partial }{\partial \eta }{\cal P}_{\text{s}}=-\overline{\phi }_{0\text{s}}\gamma
^{* \prime}(\overline{\phi }_{0\text{s}})\eta {\cal P}_{\text{s}},  \label{b21}
\end{equation}%
whose solution is 
\begin{equation}
{\cal P}_{\text{s}}\left( \eta\right) =\frac{1}{\sqrt{2B(\phib_{0\text{s}}){\pi }}}e^{-\eta
^{2}/2B(\phib_{0\text{s}})},
\quad B\left( \overline{\phi }_{0\text{s}}\right) =
\frac{1}{\gamma^{* \prime}(\overline{\phi }_{0\text{s}})}
\left[ \frac{2}{3}n_{1}^{* }(%
\overline{\phi }_{0\text{s}})+\frac{4}{5}\overline{\phi }_{0\text{s}}n_{2}^{* }(%
\overline{\phi }_{0\text{s}})\right],  \label{b22}
\end{equation}%
or, in terms of $\phi $,%
\begin{equation}
{P}_{\text{s}}(\phi)=\frac{1}{\sqrt{2B(\phib_{0\text{s}}){h\pi }}}
e^{-\left( \phi -\overline{\phi }_{0\text{s}}\right)
^{2}/2B(\phib_{0\text{s}})h}.  \label{b23}
\end{equation}

In the normal phase $\overline{\phi }_{0\text{s}}=0$ and the change of variables in
(\ref{b17}) becomes $\phi =h^{p}\eta $. A non-trivial equation for ${\cal P}
$ is obtained with $p=1$, 
\begin{equation}
\partial _{s}{\cal P}\left( \eta ,s,0\right) \rightarrow \frac{\partial }{%
\partial \eta }\left[ \left( 1-\xi ^{* }\right) \eta -\left(
1-\frac{2}{3}\frac{\partial }{\partial \eta }\eta \right) n_{1}^{* }(0)\right]
{\cal P}\left( \eta ,s,0\right) ,\quad \xi ^{* }<1.  \label{b25}
\end{equation}%
The average value of $\eta $ now obeys the equation%
\begin{equation}
\left[ \partial _{s}+\left( 1-\xi ^{* }\right) \right] \overline{\eta }%
\left( s\right) =n_{1}^{* }(0) . \label{b25a}
\end{equation}%
The stationary \ solution to this equation is 
\beq
\overline{\eta }_{\text{s}}=\frac{n_{1}^{* }(0)}{1-\xi ^{* }},
\eeq
which gives the leading finite contribution to $\overline{\phi }_{\text{s}}$ 
as $h\rightarrow 0$. The stationary distribution function is  
\begin{equation}
{\cal P}_{\text{s}}\left( \eta\right) =\frac{3}{\overline{\eta }_{\text{s}}}\left( \frac{%
3\eta }{2\overline{\eta }_{\text{s}}\pi }\right) ^{1/2}e^{-3\eta /2
\overline{\eta }_{\text{s}}},  \label{b26}
\end{equation}%
and the corresponding distribution in terms of $\phi $ is%
\begin{equation}
{P}_{\text{s}}(\phi )=\frac{3}{h\overline{\eta }_{\text{s}}}\left( \frac{3\phi }{2h\overline{%
\eta }_{\text{s}}\pi }\right) ^{1/2}e^{-3\phi /2h\overline{\eta }_{\text{s}}}.  \label{b27}
\end{equation}


\begin{references}
\bibitem{note2} Following the thermodynamic analogy, the critical exponents $%
\hat{\delta}$, $\hat{\beta}$, $\hat{\gamma}$, and $\hat{\alpha}$ are defined
as $\phib_{\text{s}} (\xi^* =1,h)\sim h^{1/\hat{\delta}}$, $\phib_{\text{s}} (\xi^* >1,h=0)\sim
(\xi^* -1)^{\hat{\beta}}$, $\chi(\xi^* ,h=0)\sim |\xi^* -1|^{-\hat{%
\gamma}}$, $C_{h}(\xi^* ,h=0)\sim |\xi^* -1|^{-\hat{\alpha}}$.
\bibitem{note1}Note
that the friction coefficient $\overline{\gamma}^*(\phib)=(1+\phib)^{1/2}$ and
noise coefficient $\overline{n}(\phib)=(1+\phib)^{3/2}$ introduced in the
phenomenological description of Section \ref{sec2} are
$\overline{\gamma}^*(\phib)=\phib^{-1}\langle \phi \gamma^*(\phi)\rangle$
and $\overline{n}(\phib)=\langle n_1^*(\phi)\rangle$, where the averages
are taken with the maximum entropy ensemble.
\end{references}
\end{document}